\documentclass[reprint, aps, pra, superscriptaddress, citeautoscript, floatfix]{revtex4-2}
\usepackage{amsmath}  
\usepackage{amssymb}  
\usepackage{bm}  
\usepackage{braket}  
\usepackage{dcolumn}  
\usepackage[colorlinks=true, citecolor={blue!80!black}, urlcolor={blue!50!black}, linkcolor = {blue!80!black}]{hyperref}  
\usepackage{enumitem}
\usepackage{float}  
\usepackage{gensymb}
\usepackage{graphicx}  
\usepackage{mathtools}  
\usepackage{upgreek}  
\usepackage{siunitx}  
\usepackage{tabularx}
\usepackage{array}
\usepackage{tikz}  
\usepackage{url}  
\usepackage{verbatim}  

\DeclareSIUnit\torr{torr}

\newcommand{\fulllabel}{\mathrm{full}}
\newcommand{\linlabel}{\mathrm{lin}}
\newcommand{\nllabel}{\mathrm{nl}}
\newcommand{\dislabel}{\mathrm{dis}}
\newcommand{\lumlabel}{\mathrm{lum}}

\usepackage[export]{adjustbox}

\begin{document} 
\date{\today} 

\title{Energy participation ratio analysis for very anharmonic superconducting circuits}

\author{Figen Yilmaz}
\affiliation{QuTech and Kavli Institute of Nanoscience, Delft University of Technology, Delft, The Netherlands}
\thanks{f.yilmaz@tudelft.nl}
\author{Siddharth Singh}
\affiliation{QuTech and Kavli Institute of Nanoscience, Delft University of Technology, Delft, The Netherlands}
\author{Martijn F.S. Zwanenburg}
\affiliation{QuTech and Kavli Institute of Nanoscience, Delft University of Technology, Delft, The Netherlands}
\author{Jinlun~Hu}
\affiliation{QuTech and Kavli Institute of Nanoscience, Delft University of Technology, Delft, The Netherlands}
\author{Taryn V. Stefanski}
\affiliation{QuTech and Kavli Institute of Nanoscience, Delft University of Technology, Delft, The Netherlands}
\affiliation{Quantum Engineering Centre for Doctoral Training, University of Bristol, Bristol, England}
\author{Christian Kraglund Andersen}
\affiliation{QuTech and Kavli Institute of Nanoscience, Delft University of Technology, Delft, The Netherlands}


\begin{abstract}
Superconducting circuits are being employed for large-scale quantum devices, and a pertinent challenge is to perform accurate numerical simulations of device parameters. One of the most advanced methods for analyzing superconducting circuit designs is the energy participation ratio (EPR) method, which constructs quantum Hamiltonians based on the energy distribution extracted from classical electromagnetic simulations. In the EPR approach, we extract linear terms from finite element simulations and add nonlinear terms using the energy participation ratio extracted from the classical simulations. However, the EPR method relies on a low-order expansion of nonlinear terms, which is prohibitive for accurately describing highly anharmonic circuits. An example of such a circuit is the fluxonium qubit, which has recently attracted increasing attention due to its high lifetimes and low error rates. In this work, we extend the EPR approach to effectively address highly nonlinear superconducting circuits, and, as a proof of concept, we apply our approach to a fluxonium qubit. Specifically, we design, fabricate, and experimentally measure a fluxonium qubit coupled to a readout resonator. We compare the measured frequencies of both the qubit and the resonator to those extracted from the EPR analysis, and we find an excellent agreement. Furthermore, we compare the dispersive shift as a function of external flux obtained from experiments with our EPR analysis and a simpler lumped element model. Our findings reveal that the EPR results closely align with the experimental data, providing more accurate estimations compared to the simplified lumped element simulations.
\end{abstract}

\maketitle
\section{Introduction}

Superconducting qubits are widely regarded as one of the leading platforms for implementing large-scale quantum processors~\cite{arute2019quantum, wu2021strong, krinner2022realizing, zhao2022realization, zhu2022quantum,  kim2023evidence, google2023suppressing, acharya2024quantum, ali2024reducing}. Superconducting circuits typically consist of linear components such as capacitors and inductors as well as nonlinear elements formed by Josephson junctions. To successfully build large-scale quantum devices, we must have a precise understanding of the Hamiltonian parameters of the system. However, the nonlinear energy of the Josephson junctions prohibits the use of fully classical methods, such as finite element simulation, to capture the relevant quantum interactions. 

So far, several methods have been used and improved to extract the parameters of superconducting quantum devices using a combination of classical and quantum methods. A common approach is to describe the quantum circuits using lumped element models, quasi-lumped models or continuum limits of lumped models in which the nonlinear components can be introduced in terms of discrete modes in the circuit~\cite{devoret1995quantum, blais2004cavity, bourassa2012josephson, leib2012networks, mortensen2016normal, parra2018quantum, minev2021circuit, egusquiza2022algebraic}. The challenge of these methods is to accurately translate real physical designs into effective model parameters. An alternative approach that solves this challenge is to analyze the device design with a black-box approach using classical finite-element simulations of the device as a basis for the quantization of the device~\cite{Nigg_2012, solgun2014blackbox, minev2021energy}. The energy participation ratio (EPR) method is an example of a method from the latter class which uses the classical energy eigenmodes of a circuit as a basis for building the quantum Hamiltonian. The EPR method reintroduces nonlinear elements to the classical energies through a perturbative expansion which has proven to be very effective for the widely used transmon qubit~\cite{koch2007charge} which exhibits weak anharmonicity. However, highly anharmonic qubits, such as the fluxonium qubit~\cite{manucharyan2009fluxonium, nguyen2019high}, have recently gained increased traction due to their very high lifetimes and gate fidelities~\cite{ding2023high, somoroff2023millisecond, zhang2024tunable, wang2024achieving}. Modelling these circuits using black-box methods is an open question and requires further advancements. 

In this work, we address the specific challenges of extracting Hamiltonian parameters for highly anharmonic superconducting quantum circuits. In particular, we present our implementation for fluxonium qubits. As proof of concept, we design, fabricate and measure a fluxonium qubit and we compare the results of our extended EPR method with the experimental data.
We demonstrate that we see a great alignment between the EPR simulations and the experimentally measured resonator and qubit frequencies. Moreover, the nonlinearity of the qubit also gives rise to a nonlinear coupling between the resonator and the qubit in the form of a dispersive shift. 
Therefore, we also analyze the dispersive shift as predicted by the extended EPR method and we again demonstrate great alignment with the experimentally measured dispersive shift. 

This paper is structured as follows: Section II introduces the improved EPR method and its application to very anharmonic superconducting circuits. Section III covers the design of fluxonium qubits. Section IV presents the experimental measurements and compares them with EPR simulation results. Section V concludes with a summary of our findings and implications for future work.

\section{Energy Participation Ratio} \label{section_2}
\begin{figure}[t] 
\centering
\includegraphics[width=1.0\linewidth]{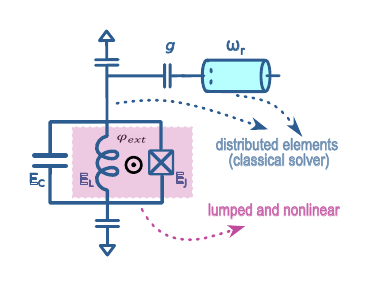}
\caption{The fluxonium consists of a linear inductor with the inductive energy $E_L$, a Josephson junction with a Josephson energy $E_J$, and a capacitor with a charging energy $E_C$. The fluxonium is capacitively coupled to a readout resonator with angular frequency $\omega_r$. The capacitive elements as well as the readout resonator (blue) can be simulated using classical finite element solvers, while the inductor and Josephson junction (pink) are represented by lumped elements.}
\label{fig:1}
\end{figure}

In this section, we briefly explain how the energy participation ratio (EPR) \cite{minev2021energy} analysis framework works and then demonstrate this method for highly anharmonic superconducting circuits~\cite{manucharyan2009fluxonium, lin2018demonstration, earnest2018realization}. The EPR method aims to provide a quantum Hamiltonian for a quantum device design through the analysis of the classical electromagnetic energy distribution. We generally design superconducting quantum devices using patterned superconducting structures, which can be combined with lumped elements. While we can treat the distinction between distributed (linear) circuits and lumped (and potentially nonlinear) circuits in broad terms for now, it is worth noting that we often encounter effective lumped elements based on Josephson junctions~\cite{josephson1962possible, josephson1965phys}; see, for example, the fluxonium qubit~\cite{manucharyan2009fluxonium} device in Fig.~\ref{fig:1}.

For electrical circuits, the linear part of a quantum Hamiltonian is equivalent to the classical Hamiltonian. Therefore, it is often convenient to split the Hamiltonian into linear and nonlinear parts:
\begin{align} \label{Hfull}
H_{\fulllabel} = H_{\linlabel} + H_{\nllabel}.
\end{align}
The linear part of the Hamiltonian contains only quadratic terms, which we split into two parts: 
\begin{align}
H_\linlabel = H_{\dislabel} + H_{\lumlabel}, \label{eq:hlin}
\end{align}
where $H_{\dislabel}$ denotes the quadratic terms representing the distributed elements of the circuit, while $H_{\lumlabel}$ represents the linear parts of the lumped structures added to the circuit. Note that the eigenmodes of the linear circuit can be fully obtained from classical finite-element methods. For now, we assume that only inductive lumped elements are added to the circuit, and we write:
\begin{align} 
H_{\lumlabel} = \frac{1}{2} \sum_j E_j (2\pi \Phi_j / \Phi_0)^2, \label{eq:hlum}
\end{align}
with $E_j$ being the inductive energy of element $j$, and $\Phi_j$ is the flux change across that element, equal to the time integral of the voltage difference across the element. Finally, $\Phi_0$ is the magnetic flux quantum. 

With the classical and linear terms set up, we can start building the quantum Hamiltonian. First, we can readily find the normal modes of the classical Hamiltonian and, thus, write the quantum Hamiltonian as a sum of quantum harmonic oscillators. The frequencies of each harmonic oscillator mode correspond to the classically solved normal mode frequencies. In other words, we can write the linear part of the Hamiltonian as:
\begin{align} 
H_\linlabel = \sum_m \hbar\omega_m a^{\dagger}_m a_m, \label{Hlin} 
\end{align}
where $\omega_m$ and $a_m$ are the angular frequency and the annihilation operator for mode $m$, respectively. Note that this Hamiltonian describes both the distributed elements $H_{\dislabel}$ and the lumped element terms in Eq.~\eqref{eq:hlin}.

Physically, the lumped inductive elements may include Josephson junctions, arrays of Josephson junctions~\cite{manucharyan2009fluxonium, masluk2012microwave, puertas2019tunable}, or kinetic inductors~\cite{hazard2019nanowire, niepce2019high, grunhaupt2019granular, rieger2023granular}, and therefore exhibit some degree of nonlinearity. We capture this nonlinearity with the Hamiltonian:
\begin{align} \label{H_nl} 
H_{\nllabel} \equiv \sum_{j=1}^J \sum_{p=3}^{\infty} E_j c_{jp} (2\pi \Phi_j/\Phi_0)^p ,
\end{align}
where $c_{jp}$ represents the coefficient of the $p$th order term in the expansion of the nonlinearity associated with element $j$. The total number of nonlinear elements is denoted with $J$. 

Now, we can in principle combine the linear Hamiltonian with the nonlinear terms and we have the full quantum Hamiltonian readily at hand. However, in the process, we need to address several challenges. The first challenge in solving the full system is that the nonlinear part $H_{\nllabel}$ can mix all modes to high orders. As a convention~\cite{minev2021energy}, the expansion of the nonlinear term is often truncated to a fixed order \textit{N}, resulting in the nonlinear Hamiltonian:
\begin{align} \label{TaylorHnl}
    H_{\nllabel} &= \sum_{j=1}^J E_j (c_{j3}\varphi_j^3 + c_{j4}\varphi_j^4 + ... )\\
    &\approx \sum_{j=1}^J E_j \sum_{p=3}^N c_{jp} \varphi_j^p,
\end{align}
with $\varphi_j=2\pi \Phi_j/\Phi_0$ introduced here for brevity. However, for very anharmonic systems, the truncation to a fixed number of nonlinear terms leads to significant errors in the calculated energies. As an example of a system that is noteworthy~\cite{ding2023high, zhang2024tunable, stefanski2024flux, lin202424} and highly nonlinear is the system that we consider in Fig.~\ref{fig:1}, which depicts a fluxonium qubit capacitively coupled to a readout resonator with a coupling strength $g$. Here, the nonlinearity arises from a single nonlinear lumped element. Therefore, we simply express the nonlinear Hamiltonian specifically for a single Josephson junction:
\begin{align}
H_{\nllabel} = -E_j[\cos(\varphi_j)+\varphi^2_j/2] \label{Hnl_j},
\end{align}
with no further approximations at this point and where $\varphi_j$ coincides with the phase difference across the junction. The next challenge appears since we need to define the quantum operator $\varphi_j$. We can use the classically solved eigenmodes as a basis to express this operator as:
\begin{align} \label{varphi_j}
    \varphi_j = \sum_{m=1}^M\varphi_{mj} (a_m^{\dagger} + a_m),
\end{align}
where $\varphi_{mj}$ is the quantum zero-point fluctuation (ZPF) of the junction phase in mode $m$.

Currently, the challenge remains to find the ZPF for each mode in the junction $\varphi_{mj}$, which we cannot directly extract from a classical simulation. However, it turns out that introducing the energy participation ratio (EPR) is useful in this context. We define the EPR as:
\begin{align} \label{p_mj}
    p_{mj} &\equiv \frac{\text{Inductive energy stored in junction } j}{ \text{Total inductive energy stored in mode } m}\\
   &= \frac{\bra{\psi_m} \frac{1}{2} E_j\varphi_j^2\ket{\psi_m}}{\bra{\psi_m} \frac{1}{2} H_{\linlabel}\ket{\psi_m}} \label{p_mj2}
\end{align}
where $\psi_m$ is a quantum state with one Fock excitation in mode $m$ while the remaining modes are in the vacuum state. In the EPR method, finite-element simulations provide us with the inductive energy in the junction, while the total inductive energy is always equal to $ \hbar \omega_m/2 $. 

Using Eq.~\eqref{p_mj2}, we calculate the energy participation from the quantum Hamiltonian, which should match the classically calculated participation ratio. Now, calculations are reduced such that we compute the zero point fluctuations ($\varphi_{mj}$): 
\begin{align} \label{pmj}
    \varphi_{mj}^2 = p_{mj} \frac{\hbar \omega_m}{2E_j}.
\end{align}
Applying this expression to calculate the ZPF for each mode, we can find the nonlinear quantum Hamiltonian using Eq. \eqref{Hnl_j} and Eq. \eqref{varphi_j}. Thus, we reach the full Hamiltonian at Eq.~\ref{Hfull}.

In the example presented in Fig.~\ref{fig:1}, we consider the case of a fluxonium qubit and a readout resonator, so we are interested in only two modes. We can identify the two modes from their differences in energy participation in the junction. Specifically, we refer to the modes with the highest EPR in the junction as the qubit mode, while the other mode with low EPR is the resonator mode. 

To highlight the importance of treating the nonlinear element without truncation, we can consider the simple Taylor expansion in Eq.~\eqref{TaylorHnl} and expand to the fourth order. When we only consider two modes $q$ and $r$, the phase operator across the junction reads
\begin{align}
\varphi_j = \varphi_{qj} (a_q^\dagger + a_q) + \varphi_{rj} (a_r^\dagger + a_r).
\end{align}
Combining the linear Hamiltonian with the fourth-order terms, we arrive at the Hamiltonian:
\begin{align}
H &= \hbar\omega_q a_q^\dagger a_q + \hbar\omega_r a_r^\dagger a_r \\ \nonumber
&- \frac{E_j}{24} [ \varphi_{qj} (a_q^\dagger + a_q) + \varphi_{rj} (a_r^\dagger + a_r)]^4.
\end{align}
By collecting the terms, we find a dispersive shift of 
\begin{align}
2\chi = E_j \varphi_{qj}^2 \varphi_{rj}^2 /12.
\end{align}
This approach is very effective for, e.g., transmon qubits \cite{koch2007charge} due to their relatively weak anharmonicity. However, as discussed above, we are focusing on fluxonium qubits, which have large anharmonicity. Concretely, we therefore implement the exact cosine through the matrix exponential of the junction operators as we detail now. For the circuit shown in Fig.~\ref{fig:1}, the nonlinearity comes, as mentioned before, from a single Josephson junction. Additionally, an external magnetic flux is threaded through the loop formed by the Josephson junction and the linear inductor. The classical simulations cannot capture the (nonlinear) effects of the external flux, thus we include the external flux into the nonlinear term associated with the Josephson junction. Therefore, as also implemented in Ref.~\cite{zhu2013circuit} for lumped circuits, the exact nonlinear term can be written as
\begin{align}
H_{\nllabel} &=-E_j [\cos(\varphi_j - \varphi_{ext})+\varphi^2_j/2] \\
&=-\frac{E_j}{2}\Big( \exp \{i [ \varphi_{qj} (a_q^\dagger + a_q) + \varphi_{rj} (a_r^\dagger + a_r)] \}e^{-i\varphi_{ext}} \nonumber \\ &\quad\quad\quad\quad + \text{h.c.}\Big) -\frac{E_j}{2}\varphi^2_j
\label{Hnl_fx}
\end{align}  
where $\varphi_{ext}=2\pi \Phi_{ext}/\Phi_0$ is the external phase bias arising from the external flux $\Phi_{ext}$ and $\varphi_{j}$ is the Josephson phase operator. We define the harmonic oscillator basis for the eigenmodes such that we express $a_q$ as a matrix in the Fock basis and then the exponential term above refers to the matrix exponential in the Fock basis of the eigenmodes. For our circuit of interest, there is no analytical expression for the dispersive shift~\cite{zhu2013circuit}. Instead, we need to diagonalize the full Hamiltonian and calculate the dispersive shift from the eigenvalues. With the eigenenergies numerically found, we can identify the dispersive shift as 
\begin{align}
2\chi = (\omega_{\ket{1,1}} - \omega_{\ket{1,0}}) - (\omega_{\ket{0,1}} - \omega_{\ket{0,0}}),
\end{align}
where $\omega_{\ket{q,r}}$ is the angular frequency for the eigenstate $\ket{q,r}$ corresponding to $q$ and $r$ excitations in the qubit and the resonator, respectively. Note that the states $\ket{q,r}$ are eigenstates of the coupled systems and not necessarily Fock states. Numerically, we first identify the resonator occupation by finding the states with maximum overlap to the state $(a_r^\dagger)^n\ket{0,0}/\sqrt{n!}$, where $\ket{0,0}$ is the readily found ground state of the system. Once the resonator excitations are identified, we can assign the qubit excitations in ascending order. 

Ultimately, through the EPR method, we extracted both the linear and nonlinear terms of the system's Hamiltonian, thereby identifying all of the device’s Hamiltonian parameters.

\section{Fluxonium Qubit Design} \label{section_3}
\begin{figure}[t] 
\includegraphics[width=1.0\linewidth]{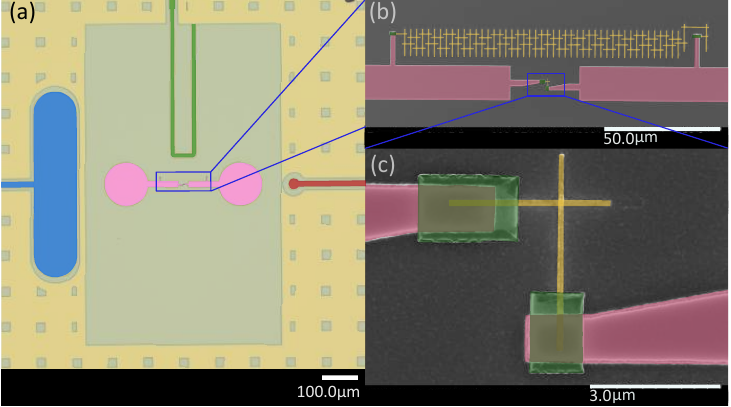} 
\caption{(a) False-coloured optical image of a fluxonium device. The pink pads represent the two superconducting islands of the fluxonium qubit, the blue pad indicates the readout pad connected to a co-planar waveguide resonator, and the green and red lines denote the flux bias and charge lines of the fluxonium qubit, respectively. The yellow area represents the ground plane of the device, while the grey area corresponds to the silicon substrate. (b) A scanning electron microscope (SEM) image with the Al junctions is shown in yellow, and Al contacts are shown in green. The central Josephson junction is located at the bottom and at the top is an array of 100 JJs which implements a large inductance. (c) The SEM image of the central Josephson junction is shown in yellow and, as in (b), the green pads are Al contacts.}
\label{fig:2}
\end{figure}

As discussed in the previous section, our extended approach to performing the EPR analysis is expected to be relevant for the prediction of the frequencies of a highly anharmonic system as well as the prediction of nonlinear interactions with such a system. Therefore, we aim to experimentally verify our approach using a circuit consisting of a fluxonium qubit capacitively coupled to a coplanar waveguide resonator (see Fig.~\ref{fig:2}). 

The fluxonium qubit is defined by three energy parameters: the Josephson junction energy, $E_J =~(\Phi_0/2e)^2/L_J$, where $L_J$ is the inductance of the Josephson junction. The inductive energy of the inductor is $E_L=~(\Phi_0/2e)^2/L$, where $L$ is the inductance. Lastly, the charging energy is defined by $E_C=e^2/(2C)$, where $C$ is the total capacitance of the qubit. We designed a floating fluxonium qubit where two circular-shaped pads provide the capacitance energy ($E_C$), the Josephson energy ($E_J$) comes from the single Josephson tunnel junction, and the inductive energy ($E_L$) is achieved with a hundred Josephson junctions in series (see Fig.~\ref{fig:2}). With this design, there is an inductive loop between the small Josephson junction and the Josephson junction array. The green line represents the external flux line carrying a DC current, which generates a magnetic field that penetrates the loop of the fluxonium qubit. This magnetic field threads the loop, setting the flux to $\Phi_{ext}$. We define the external flux as $\varphi_{ext} = 2\pi \Phi_{ext}/\Phi_0$, where $\Phi_0 = h/2e$ is the magnetic flux quantum.

The fluxonium that we study in this work, as shown in Fig.~\ref{fig:2}(a), is fabricated using a base-metalisation of NbTiN on a silicon substrate. The capacitive and co-planar waveguide structures are defined through selective etching of the NbTiN layer. To facilitate flux trapping, small squares are added to the ground plane, acting as flux-trapping holes~\cite{chiaro2016dielectric}. Both the Josephson junction array and the small Josephson junction, shown in Fig.~\ref{fig:2}(b) and (c), are fabricated using a Manhattan-style layout (yellow) during the same Al deposition process~\cite{potts2001novel}. Each Josephson junction in the array has an area of $(410\,\text{nm})^2$, and the small Josephson junction has an area of $(120\,\text{nm})^2$, see also Fig.~\ref{fig:2}(c). Galvanic contact between the junction electrodes and the NbTiN is ensured using Al patches, see Fig.~\ref{fig:2}(b) and (c) \cite{dunsworth2017characterization}. Further detailed fabrication steps are outlined in Appendix~\ref{Appendix_B}. 

We utilize the \texttt{IBM QISKIT METAL} open-source framework and library to design and conduct finite element electromagnetic simulations for our quantum devices~\cite{Qiskit_Metal}. Using \texttt{QISKIT METAL}, the designed fluxonium is rendered into \texttt{ANSYS HFSS} which performs the eigenmode simulations of the design. The results of the simulations are then automatically processed using the extended EPR method that we presented in Sec.~\ref{section_2}. The charging energy of the fluxonium is controlled by the size of the capacitor pads and the distance to the ground plane. More generally, we find the charging energy from the full capacitance network which can similarly be extracted from the finite element simulations using \texttt{ANSYS Q3D}. In particular, we fix the design of the fluxonium such that $E_C/(2\pi) = 0.943\,\text{GHz}$ based on a lumped element model. Moreover, as we discuss further below, the capacitance matrix can also be used to extract the coupling strength $g$ between the fluxonium and the readout resonator, see also Appendix \ref{Appendix_A}. Note that the eigenmode simulations include lumped inductive elements that represent both the Josephson junctions and the junction array. However, since the simulations are classical simulations, only the linear part of these elements are added to the simulation. The nonlinear parts are, as discussed above, added through the EPR analysis. In the numerical simulations, we have also added a lumped capacitive element to represent the capacitance of the Josephson junction. Conventionally, the junction capacitance has often been omitted in EPR analyses of superconducting circuits since most previous studies have focused on transmon circuits. Transmon qubits are characterized by their small charging energy corresponding to a large shunting capacitor, thus, the additional capacitance of the Josephson junction is a small additional contribution. In contrast, for fluxonium qubits, the typical capacitance associated with the qubit mode is $10\,\text{fF}$ or smaller, hence, the junction capacitance plays a larger role. Specifically, we include a junction capacitance ($C_J$) of
\begin{align}
   C_J = 50 \pm 12 fF/\mu m^2
   \label{C_J}
\end{align}
based on the measurements in Ref.~\cite{deppe2004determination}. For our case, we have a Josephson junction with an area of $(120\,\text{nm})^2$ yielding a capacitance value of $0.72\,\text{fF}$.

\section{Experimental Results} \label{Section_4}

\begin{figure}[t] 
    \includegraphics[width=1.0\linewidth]{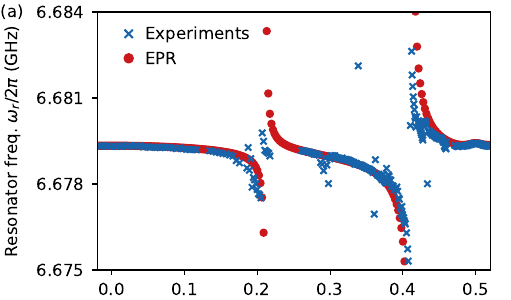}
    \includegraphics[width=1.0\linewidth]{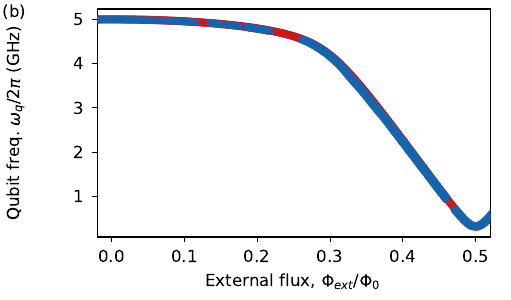} 
\caption{(a) Resonator frequency measured using spectroscopy (blue) and from the extended EPR simulations (red) as a function of the externally applied flux. (b) The qubit frequency is measured by applying a spectroscopy tone to the qubit drive line while applying a probe signal at the resonator frequency (blue). Additionally, we show the qubit frequency calculated from the extended EPR analysis (red).}
\label{fig:3}
\end{figure}

To test the applicability of the EPR analysis for the fluxonium qubit design presented in the previous section, we measure the device in a cryogenic setup (see Appendix~\ref{Appendix_C} for details of the setup). We extract the resonance frequency of the readout resonator by fitting a Lorentzian response to data obtained from spectroscopy, see blue points in Fig.~\ref{fig:3}(a). By changing the bias current in the flux line of the fluxonium, we can tune the fluxonium and, consequently, the resonator, allowing us to extract the resonator frequency as a function of the external flux.
We notice that at certain flux points, the resonator frequency exhibits transverse avoided-crossing features, which arise from the interactions between the resonator and the higher excited states of the fluxonium~\cite{zhu2013circuit, stefanski2024flux}.

Next, we perform two-tone spectroscopy to characterize the fluxonium frequency, see Fig.~\ref{fig:3}(b). By continuously driving a spectroscopy tone through the charge line of the fluxonium and simultaneously sending a readout pulse at the resonator frequency, we can detect a change in the readout signal when the spectroscopy tone is on resonance with the qubit transition. We identify the qubit frequency as the frequency with the maximal response. In contrast to the readout resonator frequency, we see that the qubit frequency shifts smoothly from around $5\,\text{GHz}$ down to around $300\,\text{MHz}$. The smooth behaviour of the qubit frequency enforces the interpretation that the avoided crossings seen in the resonator spectroscopy arise from higher states of the fluxonium. 

Lastly, we compare the measured frequencies with those predicted by the EPR analysis. To obtain the EPR results, we have three unknowns that need experimental feedback: (i) the precise resonator frequency, (ii) the Josephson energy of the fluxonium, and (iii) the inductive energy of the junction array. Starting with the resonator, its frequency is, in principle, fully defined by the geometry. However, the device is fabricated from NbTiN, which exhibits significant kinetic inductance~\cite{kroll2019magnetic}, causing a slight shift in the resonator frequency. Therefore, we offset the precise value so that the simulated frequency matches the measured frequency at zero external flux. Similarly, after analyzing the measurement data, we adjust the junction's Josephson energy and the array's inductive energy in the simulation to match the qubit frequency at zero external flux and half a magnetic flux quantum. We determine the values of $E_J/(2\pi) = 4.028\,\text{GHz}$ and $E_L/(2\pi) = 0.775\,\text{GHz}$. With these energy parameters fixed, we proceed to perform the EPR analysis as a function of the external flux, observing excellent agreement between the simulated and measured qubit frequencies, see Fig.~\ref{fig:3}. Remarkably, despite initially fixing the resonator frequency to match a single point, we successfully reproduce the flux dependence of the resonator frequency, including the avoided crossings with the higher excited states of the fluxonium, see red points in Fig.~\ref{fig:3}(b). During EPR simulations, we truncate the Hilbert space of each eigenmode to 30 Fock states.

While it is perhaps not too surprising that we can model the qubit and resonator frequencies given that we fitted the Josephson energy and the inductive energy, we note that our model can also predict higher-order coupling terms. For example, the dispersive shift is an important parameter in the context of superconducting circuits that quantifies the mutual resonance shift between the qubit and the resonator. Due to the limited selection rules of the fluxonium, higher-order qubit levels must be considered, as they have a non-negligible impact on the dispersive shift~\cite{zhu2013circuit}. Each higher-order level contributes a term to the total $\chi$, which depends on the coupling strength $g$, the charge matrix element, and the detuning from the resonator. 

\begin{figure}[t] 
\includegraphics[width=1.0\linewidth]{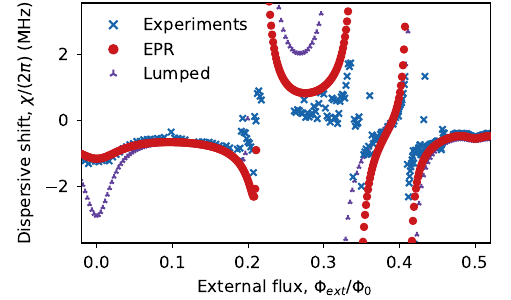}
\caption{Dispersive shift, denoted as $\chi$. Here, $\chi$ is extracted as half of the difference between resonator frequency with the qubit in the excited state with the qubit in the ground state. The experimentally measured dispersive shift (blue) is compared with the EPR results (red) and the results from a lumped element model (purple).}
\label{fig:4}
\end{figure}

In Fig.~\ref{fig:4}, we show the experimentally extracted dispersive shift (blue) and compare it with the EPR simulations (red). We perform resonator spectroscopy with the qubit prepared in the $\ket{0}$ and $\ket{1}$ states to measure the dispersive shift. When the qubit frequency of the fluxonium is below $1\,\text{GHz}$, we require an active reset protocol since the thermal equilibrium of the fluxonium is close to a completely mixed state. We employ flux-pulse-assisted reset similar to Ref. \cite{moskalenko2022high}, in which a flux pulse temporarily shifts the fluxonium to a higher frequency causing it to relax to the $\ket{0}$ state. We apply a $\pi$-pulse to the qubit after the reset to prepare the qubit in the $\ket{1}$ state. We fit double Lorentzian functions to both resonator traces to account for imperfect reset and decay during readout, providing the change in frequency of the resonator when the qubit is in $\ket{0}$ or $\ket{1}$, which amounts to twice the dispersive shift. Additionally, we can calculate the dispersive shift from the EPR analysis as discussed in Sec.~\ref{section_2}. It is evident that the EPR result closely matches the experimental data, see Fig.~\ref{fig:4} when tuning the external flux from zero external flux across half a magnetic flux quantum. Around 0.3 flux quanta, the discrepancy between the EPR results and the experimental data may originate from the fact that the experimental data is obscured by the nonlinearity of the resonator close to the avoided crossing with the higher fluxonium levels. Alternatively, the EPR analysis may also overestimate the coupling to the higher levels of the fluxonium.

Up to this point, all capacitive circuit parameters, such as the charging energy $E_C$ and the coupling strength $g$, have been implicitly included in the EPR simulations. Alternatively, we could perform an electrostatic simulation to extract the system's capacitances and then use a lumped model simulation to extract $E_C$ and the coupling strength $g$. To compare our extended EPR simulation with this simpler approach, we calculate $E_C$ and $g$ based on the lumped model as detailed in Appendix~\ref{Appendix_A}. We use \texttt{ANSYS Q3D} to extract the Maxwell capacitance matrix, from which we determine the fluxonium's charging energy 
$E_C$ and this matrix also allows us to determine the capacitance between the readout pad and the capacitor pads, which is crucial for computing the coupling strength $g$. From our calculations, we obtained the coupling strength $g/(2\pi)=85.15\,\text{MHz}$ and the charging energy $E_C/(2\pi)=0.943\,\text{GHz}$. We configure the \texttt{scQubit} package~\cite{chitta2022computer, groszkowski2021scqubits} to model a fluxonium coupled to a resonator and we use the $E_C$ and $g$ values derived from our lumped model calculations. Additionally, we incorporate the $C_J$ parameter as defined in Eq.~\eqref{C_J}. 

The dispersive shift as extracted from the simplified lumped model tracks the experimental data effectively, as shown by the purple points in Fig.~\ref{fig:4}. However, at zero flux quantum and at flux biases around 0.2 to 0.3 flux quanta, the EPR results align much more closely with the experiments compared to the lumped model. The lumped model seems therefore too simplified for the circuit considered here, where the main simplification is that all superconducting islands can be considered as equipotential surfaces. Instead, it appears that the extended mode distribution of the fluxonium mode plays a role in renormalizing the effective circuit parameters, such as $E_C$ and $g$, and, importantly, the EPR analysis readily captures this effective renormalization. As such, our extended EPR analysis can fully describe the nonlinear coupling of a highly anharmonic circuit.

\section{Conclusion}
In this work, we have designed a fluxonium qubit using \texttt{QISKIT METAL} and we extended the EPR analysis method such that we can account for the large anharmonicity of the fluxonium. When comparing the simulation results of the extended EPR method to experimental data, we find excellent agreement in terms of predicting the frequencies of the qubit and the readout resonator. Additionally, we addressed the issue of predicting the dispersive coupling of the fluxonium qubit with the readout resonator and we again found good agreement between the simulations and the experimental data. A key observation is that, when extracting the dispersive shift using a fully lumped element analysis, we found a lower accuracy compared to the experimentally measured dispersive shift. Thus, our results highlight the importance of high-frequency simulations in predicting the quantum properties of superconducting quantum devices. Looking ahead, the next apparent challenge is to integrate our method into larger circuits, e.g., featuring multiple coupled fluxonium qubits and potentially tunable coupling elements. This work represents a significant step forward in the development of efficient, scalable quantum computing technologies based on highly nonlinear superconducting qubits.

\section*{Acknowledgments}
F.Y. designed the fluxonium device. F.Y. and S.S. fabricated the device. F.Y. and J.H. developed the EPR analysis. M.F.S.Z. and T.V.S. set up the measurement electronics and software. M.F.S.Z and S.S. measured the experimental data. M.F.S.Z. and F.Y. analyzed the data. F.Y. wrote the manuscript with input from all co-authors, and C.K.A. supervised the work. Authors thank the following for input and preliminary work that helped enable this work: J. Parvizinejad, A. Kazmina, S. Vallés-Sanclemente, R. v.d. Boogaart, D. Thoen, M. Pita-Vidal and L. Splitthoff. The authors acknowledge financial support from the Dutch Research Council (NWO). The research was partly funded through the NWO Open Competition Science project OCENW.M.23.118. T.V.S. additionally acknowledges the support of the Engineering and Physical Sciences Research Council (EPSRC) under EP/SO23607/1.

\section{Data Availability}
The extended EPR repository is available \cite{EPR_Repo}, as is the Qiskit-Metal repository \cite{Qiskit-Metal_Repo}. All measurement data acquired during this experiment are also available \cite{Measurement_Data_Repo}.

\appendix

\section{Circuit quantization} \label{Appendix_A}
\begin{figure}[t] 
\includegraphics[width=1.0\linewidth]{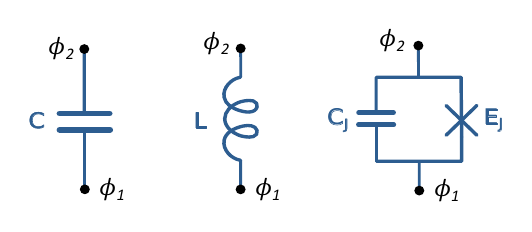}
\caption{The circuit elements of superconducting circuits. Two superconducting nodes $\phi_1$ and $\phi_2$ are connected with either a capacitor, inductor or a Josephson junction. For the Josephson junction, we have explicitly separated the Josephson capacitance and the inductive Josephson energy for clarity.}
\label{fig:app1_im1}
\end{figure}

In this appendix, we demonstrate how circuit quantization applies to superconducting circuits consisting of capacitors, inductors and Josephson junctions, see Fig.~\ref{fig:app1_im1}. To establish a quantum description of an electrical circuit, we begin with the classical Lagrangian, defined as the kinetic energy minus the potential energy ($\mathcal{L}=T-U$). For electrical circuits, the capacitance energy represents the kinetic energy, while the inductive energy serves as the system's potential energy.

We define the flux ($\phi$) and charge ($Q$) using the nodes in our quantum circuit such that
\begin{align}
\phi_n (t) = \int^{t}_{- \infty} V_n(t') dt'
\end{align}
where $V_n$ denotes node voltage at node $n$ and  
\begin{align}
Q_n (t) = \int^{t}_{- \infty} I_n(t') dt'
\end{align}
where $I_n$ denotes node current.

We use the node flux as our generalized position variable and we write the Lagrangian of each element as
\begin{align}
\mathcal{L(\phi, \dot{\phi})} = &\, \mathcal{L}_C + \mathcal{L}_L = E_{cap}(\dot{\phi}) - E_{ind}(\phi).
\end{align}
In practice, we construct this Langragian by summing up the energies of all elements in the circuit. Specifically, for a capacitor connection between two nodes, $\phi_1$ and $\phi_2$, we have the capacitive energy
\begin{align}
E_{cap}(\dot{\phi}_1, \dot{\phi}_2) =  &\, \frac{C}{2}(\dot{\phi}_1 - \dot{\phi}_2 )^2.  
\label{L_C}
\end{align}
Similarly, for an inductor connecting nodes $\phi_1$ and $\phi_2$, we have 
\begin{align}
E_{ind}({\phi}_1, \phi_2) =  &\, \frac{({\phi}_1 - {\phi}_2 )^2}{2L}. \label{L_L}
\end{align}
The next step is to introduce the Josephson junction.

The Josephson junction is characterized by a Josephson capacitance that simply takes the form of Eq.~\eqref{L_C} and an additional inductive energy
\begin{align}
E_{JJ}(\phi_1,\phi_2) = -E_J \cos\Big(2\pi \frac{\phi_1-\phi_2}{\phi_0}\Big) 
\label{L_JJ},
\end{align}
where $\phi_0$ is the magnetic flux quantum.

\begin{figure}[t] 
\includegraphics[width=1.0\linewidth]{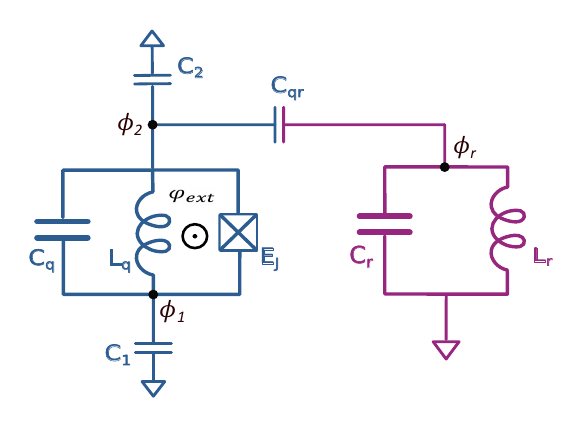}
\caption{A circuit diagram of the fluxonium qubit capacitively coupled to a resonator. The fluxonium is characterized by a Josephson energy $E_J$, an inductor with an inductance $L_q$ and a capacitor $C_q$ which includes the junction capacitance. Additionally, each fluxonium island has a capacitive connection to the ground. The resonator is here described as a lumped oscillator with capacitance $C_r$ and inductance $L_r$. The resonator and the fluxonium are capacitively coupled by a capacitance $C_{qr}$.}
\label{fig:app1-im2}
\end{figure}

After writing each element's Lagrangian, we are ready to quantize a fluxonium qubit coupled to a resonator. This approach allows us to derive the coupling strength, resonator frequency, and other relevant parameters. 
For the circuit diagram of our design, see Fig.~\ref{fig:app1-im2}, w
\begin{align}
\mathcal{L} =&\, \frac{C_q}{2}(\dot{\phi}_1 - \dot{\phi}_2 )^2 + \frac{C_1}{2} \dot{\phi}_1^2 + \frac{C_2}{2} \dot{\phi}_2^2 + \frac{C_r}{2}\dot{\phi}_r^2 \nonumber\\& + \frac{C_{qr}}{2}(\dot{\phi}_2 - \dot{\phi}_r)^2 + E_J\cos(2\pi \frac{\phi_1 - \phi_2}{\phi_0}) \label{eq:lagrangian}\\
& -\frac{1}{2L_q}(\phi_1-\phi_2)^2 - \frac{1}{2L_r}\phi_r^2. \nonumber
\end{align}
For convenience, we can write the capacitance part in matrix form such that the Lagrangian reads
\begin{align}
\mathcal{L} =&\, \frac{1}{2}\vec{\dot{\phi}}^T \mathbb{C} \vec{\dot{\phi}} + E_J\cos(2\pi \frac{\phi_1 - \phi_2}{\phi_0})\nonumber\\
& -\frac{1}{2L_q}(\phi_1-\phi_2)^2 - \frac{1}{2L_r}\phi_r^2,
\end{align}
with the matrix
\begin{align}
\mathbb{C} = \begin{pmatrix}
C_1 + C_q & -C_q & 0 \\
-C_q & C_2 + C_q + C_{rq} & -C_{rq} \\
0 & -C_{rq} & C_r + C_{rq}
\end{pmatrix}.
\end{align}
We now make the substitution $\phi = \phi_1 - \phi_2$, $\phi_\Sigma = \phi_1 + \phi_2$ and $\phi_r = \phi_r$ which is described by the transformation matrix
\begin{align}
    \mathbb{M} = \begin{pmatrix}
        1 & -1 & 0\\
        1 & 1 & 0\\
        0 & 0 & 1
    \end{pmatrix}.
\end{align}

Using this transformation, the Lagrangian becomes
\begin{align}
\mathcal{L} =&\, \frac{1}{2}[\dot{\phi}, \dot{\phi}_\Sigma, \dot{\phi}_r] \tilde{\mathbb{C}} [\dot{\phi}, \dot{\phi}_\Sigma, \dot{\phi}_r]^T + E_J\cos(2\pi \frac{\phi}{\phi_0})\nonumber\\
& -\frac{1}{2L_q}\phi^2 - \frac{1}{2L_r}\phi_r^2,
\end{align}

where we have the transformed capacitance matrix
\begin{align}
\tilde{\mathbb{C}} = \begin{pmatrix}
\frac{C_1 + C_2 + C_{qr}}{4} + C_q & \frac{C_1 - C_2 - C_{qr}}{4} & C_{qr}/2 \\
\frac{C_1 - C_2 - C_{qr}}{4} & (C_1 + C_2 + C_{qr})/4 & -C_{qr}/2 \\
C_{qr}/2 & -C_{qr}/2 & C_r +C_{qr}
\end{pmatrix}.
\end{align}
From here, we can find the conjugate variables as
\begin{align}
\vec{q} = \tilde{\mathbb{C}} [\dot{\phi}, \dot{\phi}_\Sigma, \dot{\phi}_r]^T
\end{align}
which leads to the Hamiltonian
\begin{align}
H = \frac{1}{2}\vec{q}^T \tilde{\mathbb{C}}^{-1} \vec{q} - E_J\cos\big(2\pi \frac{\phi}{\phi_0}\big) + \frac{1}{2L_q}\phi^2 + \frac{1}{2L_r}\phi_r^2.
\end{align}
Here, the charging energies are set by the inverse capacitance matrix. When we assume $C_r \gg C_1, C_2, C_{qr}, C_q$ and we ignore the $\phi_\Sigma$ terms, we find the relevant charging terms as
\begin{align}
\tilde{\mathbb{C}}^{-1} = \begin{pmatrix}
\frac{1}{C_{\star}} & - & \frac{1}{C_{coup}} \\
 - & - & - \\
 \frac{1}{C_{coup}}& - & \frac{1}{C_r+C_{qr}}
\end{pmatrix}, 
\end{align}
with 
\begin{align} \label{eq:Cstar}
    \frac{1}{C_{\star}} = \frac{1}{C_q + \frac{C_1(C_2+C_{qr})}{C_1+C_2+C_{qr}}}
\end{align}
and  
\begin{align} \label{eq:capacitance}
    \frac{1}{C_{coup}} = \frac{-C_{qr}}{C_{\star}(C_{qr}+C_r) \frac{(C_1 + C_2 + C_{qr})}{C_1}}.
\end{align}
Thus, we finally arrive at the Hamiltonian of the system
\begin{align}
H = &\, \frac{1}{2C_{\star}}q^2 - E_J\cos\big(2\pi \frac{\phi}{\phi_0}\big) + \frac{1}{2L_q}\phi^2 \nonumber\\
&\, +\frac{1}{2(C_r + C_{qr})} q_r^2 + \frac{1}{2L_r}\phi_r^2 + \frac{1}{C_{coup}}q q_r .
\end{align}
We can introduce $E_C = e^2/(2C_{\star})$ and $E_L = \frac{\phi_0^2}{L_q (2\pi)^2}$ together with the operators $n = q/(2e)$, $\varphi = 2\pi\phi/\phi_0$, $\phi_r = \sqrt{\hbar/2\omega (C_r + C_{qr})} (a^\dagger + a)$ and $q_r = i \sqrt{\hbar \omega (C_r + C_{qr})/2}(a^\dagger - a)$ such that we get the Hamiltonian on the form
\begin{align}
H = 4E_C n^2 + \frac{E_L}{2}\varphi^2 - E_J\cos(\varphi) \nonumber \\
+ \hbar \omega_r a^\dagger a + i\hbar g (a^\dagger - a)n. \label{eq:lumpedH}
\end{align}
Here we have the resonator frequency given by
\begin{align}
\omega_r = 1/\sqrt{L_r(C_r + C_{qr})}. 
\end{align}
In Eq.~\eqref{eq:lumpedH}, we have derived the coupling $g$ assuming the resonator to be a lumped LC oscillator as shown in Fig.~\ref{fig:app1-im2}. In the design presented in the main text, our readout resonator is a $\lambda/2$ resonator with a large shunt capacitor introduced by the pad close to the qubit. This large capacitance modified the mode distribution and, thus, the effective capacitive coupling~\cite{bourassa2012josephson}. Therefore, we modify the coupling $g$ to account for these two aspects and arrive at the coupling strength 
\begin{align}
g = \frac{2e \omega_r}{C_{coup}}\sqrt{\frac{Z_0}{\pi \hbar}} u_{corr} \nonumber \\
\end{align}
where $u_{corr}$ is a correction term that accounts for the precise mode distribution of the resonator. We can calculate this factor to be $u_{corr} = \cos(\omega_r C_{pad} Z_0)$ using the resonator impedance $Z_0$ and the capacitance $C_{pad}$ that refers to the combined capacitance of the readout pad.

\section{Fabrication Steps} \label{Appendix_B}

The fabrication process for this device begins with a 4-inch high-resistivity silicon wafer (resistivity $>20$ k$\Omega$cm) from Topsil, featuring a (1-0-0) orientation and with a thickness of $525\, \si{\micro\meter}$. Before metal deposition, the wafer is cleaned in nitric acid (HNO$_3$) accompanied by sonication for 6 minutes, followed by a quick dip in deionized (DI) water and then placed in a second DI water bath at room temperature ($21\degree\text{C}$). After blow-drying with $N_2$, the wafer is cleaned with 40\% HF solution for 7 minutes. It is then rinsed twice with DI water—an initial quick dip to prevent ongoing etching from residual acid—before a final blow-dry. Immediately afterwards, an HMDS layer is spun to alleviate oxide formation before sputtering~\cite{bruno2015reducing}.

We use an AJA sputtering system with a 3-inch Nb/Ti (0.7:0.3) target to deposit the metal layer, utilizing DC magnetron sputtering. The wafer is coated with a 200 nm NbTiN film under these conditions: 50 SCMM Ar and 3.5 SCMM $N_2$ plasma at a pressure of $2.3\, \si{\milli\torr}$ and a power of  $250\, \si{\watt}$. We then spin-coat the dicing resist (S1805) at $1000\, \text{rpm}$, followed by baking at $90\degree\text{C}$ for 5 minutes. In the next step, the wafer is diced into $15\times15\,\text{mm}^2$ coupons. The $15\times15\, \text{mm}^2$ coupons are placed in N-methyl pyrrolidone (NMP), rinsed with acetone and isopropyl alcohol (IPA), and blow-dried with $N_2$. 

To define the structures in the base metalisation, an AR.P 6200.18 resist is spun at 2500 rpm and baked for 3 minutes. The resist is patterned using a 100 keV Raith E-beam lithography system, operating with a beam current of $256\, \text{nA}$ and a dose of $350\,~\mu C/cm^2$. The pattern is developed using a sequence of solutions: pentyl acetate, followed by a 1:3 mixture of MIBK and IPA, and finally pure IPA. Each step lasts for 60 seconds with gentle agitation and is conducted at room temperature. Next, the NbTiN layer is selectively etched using Reactive Ion Etching (RIE) with a two-step etch recipe. The first step involves primarily physical etching, while the second step focuses on chemical etching to produce a smoother silicon surface. The etching recipes are as follows: 
\begin{enumerate}[label=\roman*)]
    \item 13.5 SCMM SF$_6$ mixed with 4 SCMM O$_2$ at 0.010~mbar pressure with 70~$\si{\watt}$ of power.
    \item 4 SCMM SF$_6$ mixed with 15 SCMM O$_2$ at 0.080~mbar pressure with 50~$\si{\watt}$ of power.
\end{enumerate}
The coupon is then placed in $80\degree\text{C}$ NMP to strip the resist overnight. Afterwards, the cleaning process is completed using acetone and IPA. Next, the coupon undergoes a 30-second BOE (1:7) dip, followed by a quick rinse in DI water and a longer soak in DI water to conclude the process. Immediately following the cleaning, a bilayer resist stack is applied to the coupon:
\begin{enumerate}[label=\roman*)]
    \item MMA-EL8 resist, spin at 4000 rpm and bake at $185\degree\text{C}$ for 5 mins after.
    \item PMMA-950 A4 resist, spin at 2000 rpm and bake at $185\degree\text{C}$ for 5 mins after.
\end{enumerate}
We use the same E-beam lithography system for the Josephson junction structures, exposing all the junctions in a single step with a Manhattan-style pattern and an undercut profile in the bilayer stack. A beam current of $98\,\text{pA}$ with a dose of $1200\,\mu\text{C/cm}^2$ is used for junction exposure, and $129\,\text{pA}$ with a dose of $540\,\mu\text{C/cm}^2$ for the undercuts. Development is performed using a H$_2$O mixture at $6^\circ\text{C}$, followed by a 40-second BOE (1:7) clean and DI water rinses lasting 10 seconds and 60 seconds, respectively. 

The Josephson junctions are deposited using the Plassys MEB 550 system via a two-step aluminium evaporation process. The details of the Al deposition are as follows: 
\begin{enumerate}[label=\roman*)] 
    \item The chamber is pre-conditioned using deposition of Ti at a rate of 0.2 nm/s until a thickness of 20 nm is reached. 
    \item The first Al deposition occurs at a rate of 0.5 nm/s with a deposition angle of 55\degree, ~reaching approximately 30 nm thickness. 
    \item The sample undergoes 11 minutes of O$_2$ oxidation at 0\degree ~and $1.3\,\text{mbar}$. 
    \item The holder is rotated by 90\degree ~for the second Al deposition, again at a rate of 0.5 nm/s and a deposition angle of 55\degree, until approximately 110 nm thickness is achieved. 
    \item The sample undergoes another 11 minutes of O$_2$ oxidation at 0\degree ~and $1.3\,\text{mbar}$. \end{enumerate}
The deposition thickness is measured on the crystal, though it is expected to be thinner on the device surface due to the deposition angle. After deposition, the chip undergoes a lift-off process, during which it is soaked in acetone at 50°C for at least 2 hours. This is followed by an additional stripping step in $80\degree\text{C}$ NMP. Finally, the chip is rinsed in succession with acetone and IPA.

After the junction fabrication is completed, patches are applied to specific areas of the junctions to improve the galvanic connection between the electrodes and the underlying metal layer (see Fig.~\ref{fig:2}). This step follows the same lithography process used for the junctions, though undercuts are not required this time. The patches are deposited using the Plassys system, beginning with a 2-minute Argon milling step in the load lock, followed by the deposition of 150 nm of Al at a 0\degree deposition angle. The chip then undergoes a lift-off process, utilizing the same acetone and NMP stripping method.

In the final step, we apply the dicing resist (S1805) by spinning it to achieve a PCB size of $9\times9\, \text{mm}^2$. After dicing with the same tool, we strip the resist, and the device is released from the cleanroom, ready to be sent to the wire-bonder room

\begin{figure}[b] 
\includegraphics[width=1.0\linewidth]{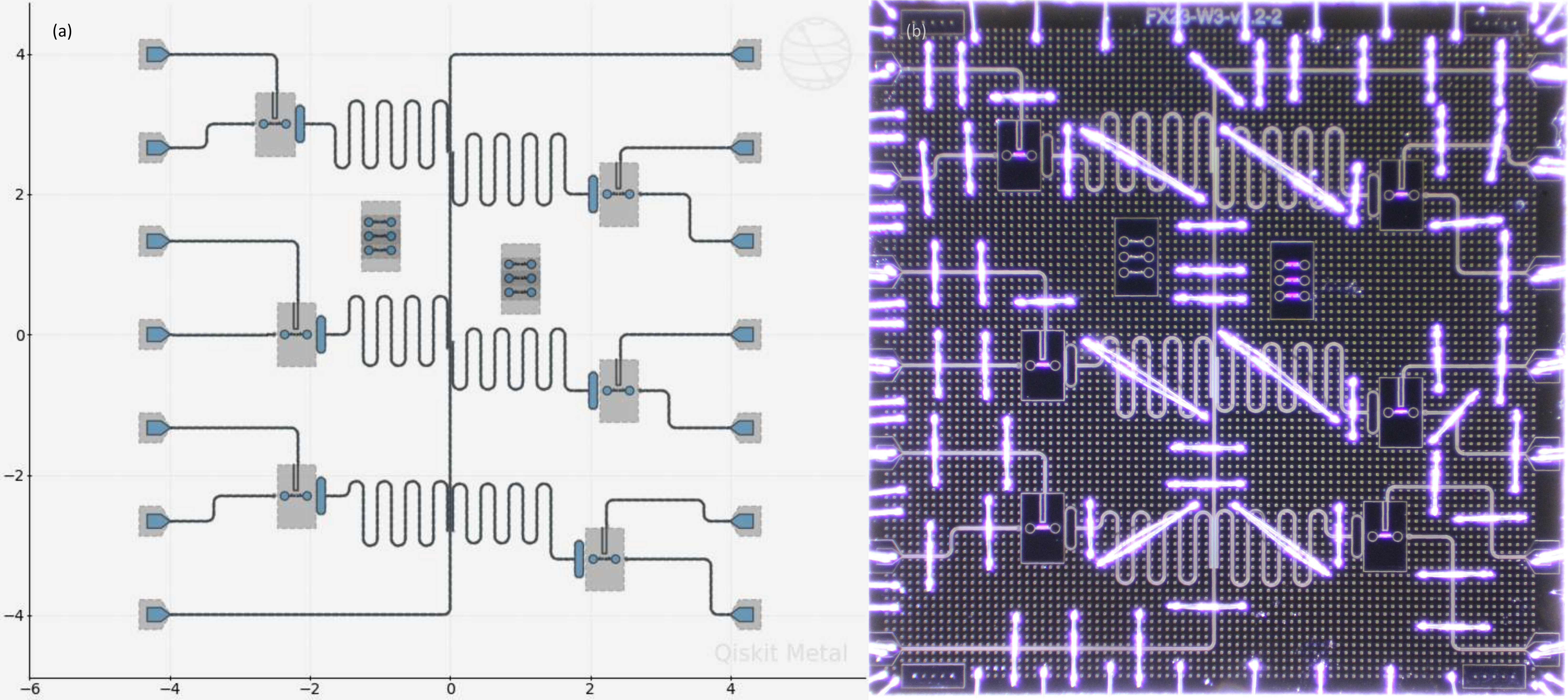}
\caption{A GUI image of the six single fluxonium qubit device (a) design by using \texttt{QISKIT METAL} (b) and the fabricated device with Al wire bonds.}
\label{fig:app_imB}
\end{figure}

In Fig.~\ref{fig:app_imB}, we display the whole circuit design (a) the device from the IBM \texttt{QISKIT METAL} GUI image and (b) the fabricated device, which includes its Al wire bonds. Some of these bonds are used for grounding, while others establish the connection between the PCB and the launch pads on the device. Additionally, we incorporate six test structures in the centre of the chip, enabling room-temperature probing of the junction resistance. Three of these structures host a single Josephson junction on the left, while the remaining three feature an array on the right.

\begin{figure} [t]
\centering
\includegraphics[width=1.0\linewidth]{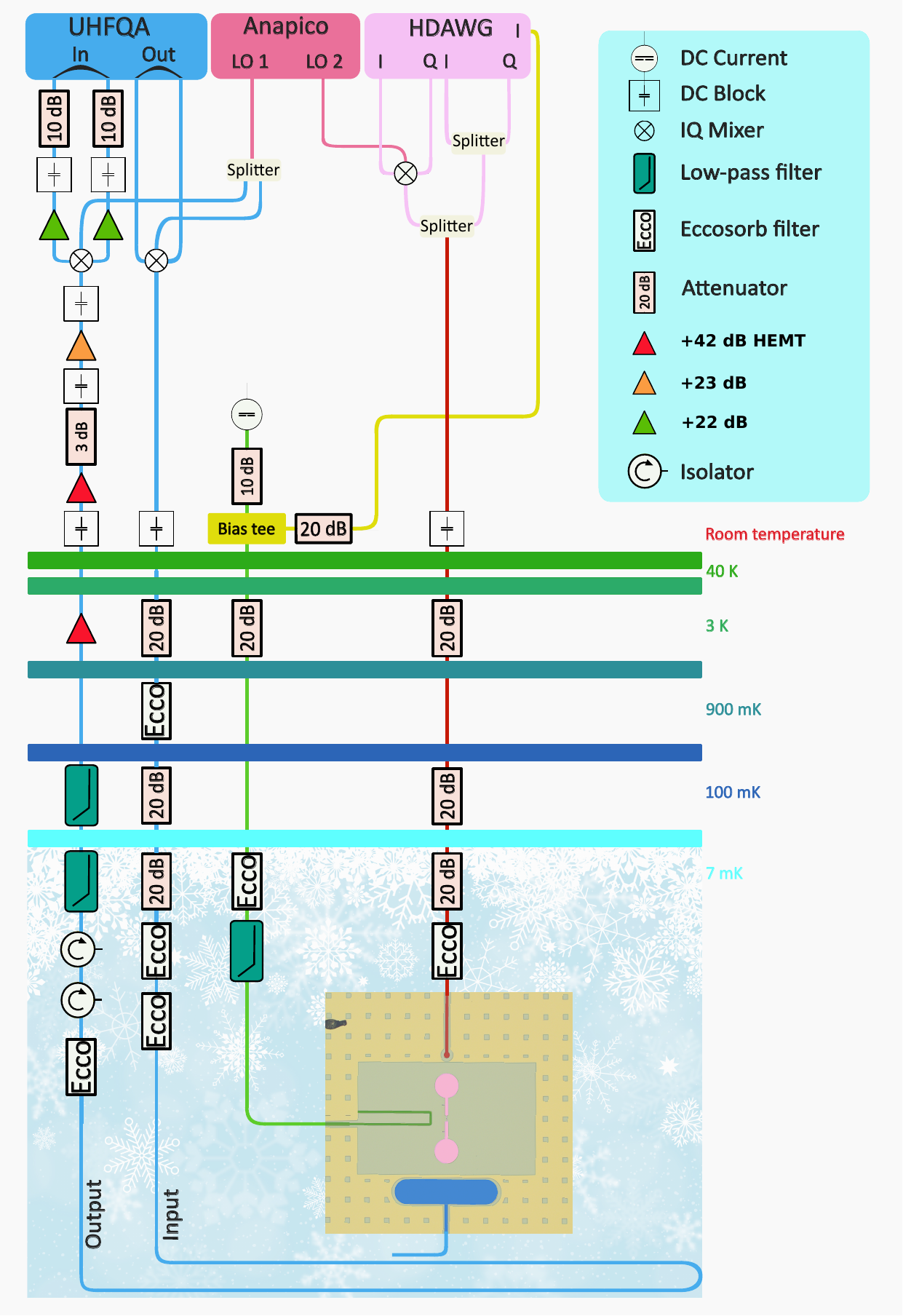} 
\caption{Wiring diagram inside the dilution refrigerator and the room temperature electronic setup, see details of the components in Table~\ref{tab:equipment} which is further discussed in the text.}
\label{fig:App_imC}
\end{figure}

\section{Measurement Setup} \label{Appendix_C}

The fabricated device is diced into $9\times9\, \text{mm}^2$ coupons, then affixed with GE varnish onto a gold-plated copper mount, and electrically connected to its Printed Circuit Board (PCB) via $25\, \si{\micro\meter}$ thick Al wire bonds using a semi-automatic F\&S Bondtec$^{\textsuperscript{\textregistered}}$ equipment. Once wire bonding is completed, the device is loaded into our dilution refrigerator. For our samples, we use BLUEFORS$^{\textsuperscript{\textregistered}}$ LD400 series, which has multiple temperature stages, as depicted in Fig.~\ref{fig:App_imC}. The samples are mounted in the mixing chamber stage, which operates at around $7\, \si{\milli\kelvin}$. For added magnetic shielding, the device is protected on both sides by Al and the PCB is encased in two Mu-metal cans.

The device used in this paper hosts six single fluxonium qubits, six flux-bias lines, six charge lines, and six $\lambda/2$ resonators, which are capacitively coupled to the transmission line, as depicted in Fig.~\ref{fig:app_imB}. However, only the wiring diagram for one qubit is shown, as we focus on a specific qubit for this work. 
\begin{table*}[t]
\centering
\begin{tabularx}{\textwidth}{>{\centering\arraybackslash}X >{\centering\arraybackslash}X}
\hline \hline
Dilution refrigerator       & BLUEFORS$^{\textsuperscript{\textregistered}}$ LD400 \\ \hline 
IQ Mixer Readout input      & Zurich Instruments$^{\textsuperscript{\textregistered}}$ HDIQ  \\ \hline 
AWG Readout output / ADC    & Zurich Instruments$^{\textsuperscript{\textregistered}}$ UHFQA \\ \hline
AWG Flux/Charge-line        & Zurich Instruments$^{\textsuperscript{\textregistered}}$ HDAWG \\ \hline
IQ Mixer Charge-line        & Marki IQ-1545LMP  \\ \hline
IQ Demodulator              & QuTech-in house \\ \hline
RF Source                   & APMS20G-4 \\ \hline
DC Source                   & QuTech-in house  \\ \hline
Bias-Tee                    & Mini-Circuits ZFBT-4R2GW+ \\ \hline
DC Block                    & Mini-Circuits BLKD-183-S+ inner-outer \\ \hline
Eccosorb IR filter          & QuTech-in house  \\ \hline
Isolator                    & LNF-ISISC4\_8A s/n 2027-17 4-8 GHz \\ \hline
Low-pass filter             & VLFX-105+   \\ \hline
Splitter                    & Mini-Circuits ZX10R-14-S+ DC-10000 MHz \\ \hline
-20 dB attenuator           & Bluefors \\ \hline
-10 dB attenuator           & MCL BW-S10 W2+ \\ \hline
+22 amplifier               & Mini-Circuits GALI-3+ \\ \hline
+42 HEMT                    & LNF-LNC4\_8C s/n 2701H \\ \hline
\end{tabularx}
\caption{List of control electronics and additional equipment used at room temperature and cryogenic stages.}
\label{tab:equipment}
\end{table*}
The room-temperature control and readout electronics are shown at the top of the drawing in Fig.~\ref{fig:App_imC}. 

The readout signal tone is generated by a Zurich Instruments UHFQA quantum analyzer and upconverted to microwave frequencies using an IQ mixer, with the local oscillator (LO 1) signal supplied by the APMS20G-4 AnaPico signal generator for the upconversion. The upconverted signal is attenuated by a series of attenuators and an Eccosorb filter before reaching the sample at base temperature. After passing through a weak Eccosorb filter, the output line includes two isolators. To filter out higher frequency components, low-pass filters are applied at $7\,\text{mK}$ and $100\,\text{mK}$. The signal is amplified at the $3\,\text{mK}$ stage by a $42\,\text{dB}$ HEMT amplifier, followed by additional $42\,\text{dB}$ and $23\,\text{dB}$ amplifiers at room temperature, with a $3\,\text{dB}$ attenuator in between to suppress standing modes. Finally, the signal is downconverted and amplified with $22\,\text{dB}$ amplifiers on both the I and Q signals. Two $10\,\text{dB}$ attenuators are used at the UHFQA input to match its dynamic range.

On the flux line, a $750\,\text{MHz}$ 8-channel High-Density Arbitrary Waveform Generator (HDAWG) generates the flux pulses, while an in-house QuTech S4g current source module supplies the DC bias current. These signals were combined with a bias-tee at room temperature. Additionally, a $20\,\text{dB}$ attenuator is included at $3\,\text{K}$, along with an Eccosorb and low-pass filters at the mixing chamber. 

We use two slightly different hardware configurations to address the qubit above and below $\omega_q/2\pi = 1\,\text{GHz}$. For frequencies above $1\,\text{GHz}$ (as shown in Fig \ref{fig:App_imC}), the HDAWG generates the \textit{IQ} signals at an intermediate frequency for upconversion. For qubit frequencies below $1\,\text{GHz}$, the drive signal is generated directly by the HDAWG. These signals are combined with a power splitter. The charge line signal is attenuated similarly to the input line, as shown in Fig.~\ref{fig:App_imC}.



\bibliography{main}

\begin{thebibliography}{53}%
\makeatletter
\providecommand \@ifxundefined [1]{%
 \@ifx{#1\undefined}
}%
\providecommand \@ifnum [1]{%
 \ifnum #1\expandafter \@firstoftwo
 \else \expandafter \@secondoftwo
 \fi
}%
\providecommand \@ifx [1]{%
 \ifx #1\expandafter \@firstoftwo
 \else \expandafter \@secondoftwo
 \fi
}%
\providecommand \natexlab [1]{#1}%
\providecommand \enquote  [1]{``#1''}%
\providecommand \bibnamefont  [1]{#1}%
\providecommand \bibfnamefont [1]{#1}%
\providecommand \citenamefont [1]{#1}%
\providecommand \href@noop [0]{\@secondoftwo}%
\providecommand \href [0]{\begingroup \@sanitize@url \@href}%
\providecommand \@href[1]{\@@startlink{#1}\@@href}%
\providecommand \@@href[1]{\endgroup#1\@@endlink}%
\providecommand \@sanitize@url [0]{\catcode `\\12\catcode `\$12\catcode
  `\&12\catcode `\#12\catcode `\^12\catcode `\_12\catcode `\%12\relax}%
\providecommand \@@startlink[1]{}%
\providecommand \@@endlink[0]{}%
\providecommand \url  [0]{\begingroup\@sanitize@url \@url }%
\providecommand \@url [1]{\endgroup\@href {#1}{\urlprefix }}%
\providecommand \urlprefix  [0]{URL }%
\providecommand \Eprint [0]{\href }%
\providecommand \doibase [0]{https://doi.org/}%
\providecommand \selectlanguage [0]{\@gobble}%
\providecommand \bibinfo  [0]{\@secondoftwo}%
\providecommand \bibfield  [0]{\@secondoftwo}%
\providecommand \translation [1]{[#1]}%
\providecommand \BibitemOpen [0]{}%
\providecommand \bibitemStop [0]{}%
\providecommand \bibitemNoStop [0]{.\EOS\space}%
\providecommand \EOS [0]{\spacefactor3000\relax}%
\providecommand \BibitemShut  [1]{\csname bibitem#1\endcsname}%
\let\auto@bib@innerbib\@empty
\bibitem [{\citenamefont {Arute}\ \emph {et~al.}(2019)\citenamefont {Arute},
  \citenamefont {Arya}, \citenamefont {Babbush}, \citenamefont {Bacon},
  \citenamefont {Bardin}, \citenamefont {Barends}, \citenamefont {Biswas},
  \citenamefont {Boixo}, \citenamefont {Brandao}, \citenamefont {Buell} \emph
  {et~al.}}]{arute2019quantum}%
  \BibitemOpen
  \bibfield  {author} {\bibinfo {author} {\bibfnamefont {F.}~\bibnamefont
  {Arute}}, \bibinfo {author} {\bibfnamefont {K.}~\bibnamefont {Arya}},
  \bibinfo {author} {\bibfnamefont {R.}~\bibnamefont {Babbush}}, \bibinfo
  {author} {\bibfnamefont {D.}~\bibnamefont {Bacon}}, \bibinfo {author}
  {\bibfnamefont {J.~C.}\ \bibnamefont {Bardin}}, \bibinfo {author}
  {\bibfnamefont {R.}~\bibnamefont {Barends}}, \bibinfo {author} {\bibfnamefont
  {R.}~\bibnamefont {Biswas}}, \bibinfo {author} {\bibfnamefont
  {S.}~\bibnamefont {Boixo}}, \bibinfo {author} {\bibfnamefont {F.~G.}\
  \bibnamefont {Brandao}}, \bibinfo {author} {\bibfnamefont {D.~A.}\
  \bibnamefont {Buell}}, \emph {et~al.},\ }\bibfield  {title} {\bibinfo {title}
  {Quantum supremacy using a programmable superconducting processor},\
  }\href@noop {} {\bibfield  {journal} {\bibinfo  {journal} {Nature}\ }\textbf
  {\bibinfo {volume} {574}},\ \bibinfo {pages} {505} (\bibinfo {year}
  {2019})}\BibitemShut {NoStop}%
\bibitem [{\citenamefont {Wu}\ \emph {et~al.}(2021)\citenamefont {Wu},
  \citenamefont {Bao}, \citenamefont {Cao}, \citenamefont {Chen}, \citenamefont
  {Chen}, \citenamefont {Chen}, \citenamefont {Chung}, \citenamefont {Deng},
  \citenamefont {Du}, \citenamefont {Fan} \emph {et~al.}}]{wu2021strong}%
  \BibitemOpen
  \bibfield  {author} {\bibinfo {author} {\bibfnamefont {Y.}~\bibnamefont
  {Wu}}, \bibinfo {author} {\bibfnamefont {W.-S.}\ \bibnamefont {Bao}},
  \bibinfo {author} {\bibfnamefont {S.}~\bibnamefont {Cao}}, \bibinfo {author}
  {\bibfnamefont {F.}~\bibnamefont {Chen}}, \bibinfo {author} {\bibfnamefont
  {M.-C.}\ \bibnamefont {Chen}}, \bibinfo {author} {\bibfnamefont
  {X.}~\bibnamefont {Chen}}, \bibinfo {author} {\bibfnamefont {T.-H.}\
  \bibnamefont {Chung}}, \bibinfo {author} {\bibfnamefont {H.}~\bibnamefont
  {Deng}}, \bibinfo {author} {\bibfnamefont {Y.}~\bibnamefont {Du}}, \bibinfo
  {author} {\bibfnamefont {D.}~\bibnamefont {Fan}}, \emph {et~al.},\ }\bibfield
   {title} {\bibinfo {title} {Strong quantum computational advantage using a
  superconducting quantum processor},\ }\href@noop {} {\bibfield  {journal}
  {\bibinfo  {journal} {Physical review letters}\ }\textbf {\bibinfo {volume}
  {127}},\ \bibinfo {pages} {180501} (\bibinfo {year} {2021})}\BibitemShut
  {NoStop}%
\bibitem [{\citenamefont {Krinner}\ \emph {et~al.}(2022)\citenamefont
  {Krinner}, \citenamefont {Lacroix}, \citenamefont {Remm}, \citenamefont
  {Di~Paolo}, \citenamefont {Genois}, \citenamefont {Leroux}, \citenamefont
  {Hellings}, \citenamefont {Lazar}, \citenamefont {Swiadek}, \citenamefont
  {Herrmann} \emph {et~al.}}]{krinner2022realizing}%
  \BibitemOpen
  \bibfield  {author} {\bibinfo {author} {\bibfnamefont {S.}~\bibnamefont
  {Krinner}}, \bibinfo {author} {\bibfnamefont {N.}~\bibnamefont {Lacroix}},
  \bibinfo {author} {\bibfnamefont {A.}~\bibnamefont {Remm}}, \bibinfo {author}
  {\bibfnamefont {A.}~\bibnamefont {Di~Paolo}}, \bibinfo {author}
  {\bibfnamefont {E.}~\bibnamefont {Genois}}, \bibinfo {author} {\bibfnamefont
  {C.}~\bibnamefont {Leroux}}, \bibinfo {author} {\bibfnamefont
  {C.}~\bibnamefont {Hellings}}, \bibinfo {author} {\bibfnamefont
  {S.}~\bibnamefont {Lazar}}, \bibinfo {author} {\bibfnamefont
  {F.}~\bibnamefont {Swiadek}}, \bibinfo {author} {\bibfnamefont
  {J.}~\bibnamefont {Herrmann}}, \emph {et~al.},\ }\bibfield  {title} {\bibinfo
  {title} {Realizing repeated quantum error correction in a distance-three
  surface code},\ }\href@noop {} {\bibfield  {journal} {\bibinfo  {journal}
  {Nature}\ }\textbf {\bibinfo {volume} {605}},\ \bibinfo {pages} {669}
  (\bibinfo {year} {2022})}\BibitemShut {NoStop}%
\bibitem [{\citenamefont {Zhao}\ \emph {et~al.}(2022)\citenamefont {Zhao},
  \citenamefont {Ye}, \citenamefont {Huang}, \citenamefont {Zhang},
  \citenamefont {Wu}, \citenamefont {Guan}, \citenamefont {Zhu}, \citenamefont
  {Wei}, \citenamefont {He}, \citenamefont {Cao} \emph
  {et~al.}}]{zhao2022realization}%
  \BibitemOpen
  \bibfield  {author} {\bibinfo {author} {\bibfnamefont {Y.}~\bibnamefont
  {Zhao}}, \bibinfo {author} {\bibfnamefont {Y.}~\bibnamefont {Ye}}, \bibinfo
  {author} {\bibfnamefont {H.-L.}\ \bibnamefont {Huang}}, \bibinfo {author}
  {\bibfnamefont {Y.}~\bibnamefont {Zhang}}, \bibinfo {author} {\bibfnamefont
  {D.}~\bibnamefont {Wu}}, \bibinfo {author} {\bibfnamefont {H.}~\bibnamefont
  {Guan}}, \bibinfo {author} {\bibfnamefont {Q.}~\bibnamefont {Zhu}}, \bibinfo
  {author} {\bibfnamefont {Z.}~\bibnamefont {Wei}}, \bibinfo {author}
  {\bibfnamefont {T.}~\bibnamefont {He}}, \bibinfo {author} {\bibfnamefont
  {S.}~\bibnamefont {Cao}}, \emph {et~al.},\ }\bibfield  {title} {\bibinfo
  {title} {Realization of an error-correcting surface code with superconducting
  qubits},\ }\href@noop {} {\bibfield  {journal} {\bibinfo  {journal} {Physical
  Review Letters}\ }\textbf {\bibinfo {volume} {129}},\ \bibinfo {pages}
  {030501} (\bibinfo {year} {2022})}\BibitemShut {NoStop}%
\bibitem [{\citenamefont {Zhu}\ \emph {et~al.}(2022)\citenamefont {Zhu},
  \citenamefont {Cao}, \citenamefont {Chen}, \citenamefont {Chen},
  \citenamefont {Chen}, \citenamefont {Chung}, \citenamefont {Deng},
  \citenamefont {Du}, \citenamefont {Fan}, \citenamefont {Gong} \emph
  {et~al.}}]{zhu2022quantum}%
  \BibitemOpen
  \bibfield  {author} {\bibinfo {author} {\bibfnamefont {Q.}~\bibnamefont
  {Zhu}}, \bibinfo {author} {\bibfnamefont {S.}~\bibnamefont {Cao}}, \bibinfo
  {author} {\bibfnamefont {F.}~\bibnamefont {Chen}}, \bibinfo {author}
  {\bibfnamefont {M.-C.}\ \bibnamefont {Chen}}, \bibinfo {author}
  {\bibfnamefont {X.}~\bibnamefont {Chen}}, \bibinfo {author} {\bibfnamefont
  {T.-H.}\ \bibnamefont {Chung}}, \bibinfo {author} {\bibfnamefont
  {H.}~\bibnamefont {Deng}}, \bibinfo {author} {\bibfnamefont {Y.}~\bibnamefont
  {Du}}, \bibinfo {author} {\bibfnamefont {D.}~\bibnamefont {Fan}}, \bibinfo
  {author} {\bibfnamefont {M.}~\bibnamefont {Gong}}, \emph {et~al.},\
  }\bibfield  {title} {\bibinfo {title} {Quantum computational advantage via
  60-qubit 24-cycle random circuit sampling},\ }\href@noop {} {\bibfield
  {journal} {\bibinfo  {journal} {Science bulletin}\ }\textbf {\bibinfo
  {volume} {67}},\ \bibinfo {pages} {240} (\bibinfo {year} {2022})}\BibitemShut
  {NoStop}%
\bibitem [{\citenamefont {Kim}\ \emph {et~al.}(2023)\citenamefont {Kim},
  \citenamefont {Eddins}, \citenamefont {Anand}, \citenamefont {Wei},
  \citenamefont {Van Den~Berg}, \citenamefont {Rosenblatt}, \citenamefont
  {Nayfeh}, \citenamefont {Wu}, \citenamefont {Zaletel}, \citenamefont {Temme}
  \emph {et~al.}}]{kim2023evidence}%
  \BibitemOpen
  \bibfield  {author} {\bibinfo {author} {\bibfnamefont {Y.}~\bibnamefont
  {Kim}}, \bibinfo {author} {\bibfnamefont {A.}~\bibnamefont {Eddins}},
  \bibinfo {author} {\bibfnamefont {S.}~\bibnamefont {Anand}}, \bibinfo
  {author} {\bibfnamefont {K.~X.}\ \bibnamefont {Wei}}, \bibinfo {author}
  {\bibfnamefont {E.}~\bibnamefont {Van Den~Berg}}, \bibinfo {author}
  {\bibfnamefont {S.}~\bibnamefont {Rosenblatt}}, \bibinfo {author}
  {\bibfnamefont {H.}~\bibnamefont {Nayfeh}}, \bibinfo {author} {\bibfnamefont
  {Y.}~\bibnamefont {Wu}}, \bibinfo {author} {\bibfnamefont {M.}~\bibnamefont
  {Zaletel}}, \bibinfo {author} {\bibfnamefont {K.}~\bibnamefont {Temme}},
  \emph {et~al.},\ }\bibfield  {title} {\bibinfo {title} {Evidence for the
  utility of quantum computing before fault tolerance},\ }\href@noop {}
  {\bibfield  {journal} {\bibinfo  {journal} {Nature}\ }\textbf {\bibinfo
  {volume} {618}},\ \bibinfo {pages} {500} (\bibinfo {year}
  {2023})}\BibitemShut {NoStop}%
\bibitem [{\citenamefont {Asfaw}\ \emph {et~al.}(2023)\citenamefont {Asfaw},
  \citenamefont {Megrant}, \citenamefont {Jones}, \citenamefont {Gidney},
  \citenamefont {Bacon}, \citenamefont {Debroy}, \citenamefont {Kafri},
  \citenamefont {Lucero}, \citenamefont {Neven}, \citenamefont {Hilton} \emph
  {et~al.}}]{google2023suppressing}%
  \BibitemOpen
  \bibfield  {author} {\bibinfo {author} {\bibfnamefont {A.}~\bibnamefont
  {Asfaw}}, \bibinfo {author} {\bibfnamefont {A.}~\bibnamefont {Megrant}},
  \bibinfo {author} {\bibfnamefont {C.}~\bibnamefont {Jones}}, \bibinfo
  {author} {\bibfnamefont {C.}~\bibnamefont {Gidney}}, \bibinfo {author}
  {\bibfnamefont {D.}~\bibnamefont {Bacon}}, \bibinfo {author} {\bibfnamefont
  {D.}~\bibnamefont {Debroy}}, \bibinfo {author} {\bibfnamefont
  {D.}~\bibnamefont {Kafri}}, \bibinfo {author} {\bibfnamefont
  {E.}~\bibnamefont {Lucero}}, \bibinfo {author} {\bibfnamefont
  {H.}~\bibnamefont {Neven}}, \bibinfo {author} {\bibfnamefont
  {J.}~\bibnamefont {Hilton}}, \emph {et~al.},\ }\bibfield  {title} {\bibinfo
  {title} {Suppressing quantum errors by scaling a surface code logical
  qubit},\ }\href@noop {} {\bibfield  {journal} {\bibinfo  {journal} {Nature}\
  }\textbf {\bibinfo {volume} {614}},\ \bibinfo {pages} {676} (\bibinfo {year}
  {2023})}\BibitemShut {NoStop}%
\bibitem [{\citenamefont {Acharya}\ \emph {et~al.}(2024)\citenamefont
  {Acharya}, \citenamefont {Aghababaie-Beni}, \citenamefont {Aleiner},
  \citenamefont {Andersen}, \citenamefont {Ansmann}, \citenamefont {Arute},
  \citenamefont {Arya}, \citenamefont {Asfaw}, \citenamefont {Astrakhantsev},
  \citenamefont {Atalaya} \emph {et~al.}}]{acharya2024quantum}%
  \BibitemOpen
  \bibfield  {author} {\bibinfo {author} {\bibfnamefont {R.}~\bibnamefont
  {Acharya}}, \bibinfo {author} {\bibfnamefont {L.}~\bibnamefont
  {Aghababaie-Beni}}, \bibinfo {author} {\bibfnamefont {I.}~\bibnamefont
  {Aleiner}}, \bibinfo {author} {\bibfnamefont {T.~I.}\ \bibnamefont
  {Andersen}}, \bibinfo {author} {\bibfnamefont {M.}~\bibnamefont {Ansmann}},
  \bibinfo {author} {\bibfnamefont {F.}~\bibnamefont {Arute}}, \bibinfo
  {author} {\bibfnamefont {K.}~\bibnamefont {Arya}}, \bibinfo {author}
  {\bibfnamefont {A.}~\bibnamefont {Asfaw}}, \bibinfo {author} {\bibfnamefont
  {N.}~\bibnamefont {Astrakhantsev}}, \bibinfo {author} {\bibfnamefont
  {J.}~\bibnamefont {Atalaya}}, \emph {et~al.},\ }\bibfield  {title} {\bibinfo
  {title} {Quantum error correction below the surface code threshold},\
  }\href@noop {} {\bibfield  {journal} {\bibinfo  {journal} {arXiv preprint
  arXiv:2408.13687}\ } (\bibinfo {year} {2024})}\BibitemShut {NoStop}%
\bibitem [{\citenamefont {Ali}\ \emph {et~al.}(2024)\citenamefont {Ali},
  \citenamefont {Marques}, \citenamefont {Crawford}, \citenamefont {Majaniemi},
  \citenamefont {Serra-Peralta}, \citenamefont {Byfield}, \citenamefont
  {Varbanov}, \citenamefont {Terhal}, \citenamefont {DiCarlo},\ and\
  \citenamefont {Campbell}}]{ali2024reducing}%
  \BibitemOpen
  \bibfield  {author} {\bibinfo {author} {\bibfnamefont {H.}~\bibnamefont
  {Ali}}, \bibinfo {author} {\bibfnamefont {J.}~\bibnamefont {Marques}},
  \bibinfo {author} {\bibfnamefont {O.}~\bibnamefont {Crawford}}, \bibinfo
  {author} {\bibfnamefont {J.}~\bibnamefont {Majaniemi}}, \bibinfo {author}
  {\bibfnamefont {M.}~\bibnamefont {Serra-Peralta}}, \bibinfo {author}
  {\bibfnamefont {D.}~\bibnamefont {Byfield}}, \bibinfo {author} {\bibfnamefont
  {B.}~\bibnamefont {Varbanov}}, \bibinfo {author} {\bibfnamefont {B.~M.}\
  \bibnamefont {Terhal}}, \bibinfo {author} {\bibfnamefont {L.}~\bibnamefont
  {DiCarlo}},\ and\ \bibinfo {author} {\bibfnamefont {E.~T.}\ \bibnamefont
  {Campbell}},\ }\bibfield  {title} {\bibinfo {title} {Reducing the error rate
  of a superconducting logical qubit using analog readout information},\
  }\href@noop {} {\bibfield  {journal} {\bibinfo  {journal} {Physical Review
  Applied}\ }\textbf {\bibinfo {volume} {22}},\ \bibinfo {pages} {044031}
  (\bibinfo {year} {2024})}\BibitemShut {NoStop}%
\bibitem [{\citenamefont {Devoret}\ \emph {et~al.}(1995)\citenamefont {Devoret}
  \emph {et~al.}}]{devoret1995quantum}%
  \BibitemOpen
  \bibfield  {author} {\bibinfo {author} {\bibfnamefont {M.~H.}\ \bibnamefont
  {Devoret}} \emph {et~al.},\ }\bibfield  {title} {\bibinfo {title} {Quantum
  fluctuations in electrical circuits},\ }\href@noop {} {\bibfield  {journal}
  {\bibinfo  {journal} {Les Houches, Session LXIII}\ }\textbf {\bibinfo
  {volume} {7}},\ \bibinfo {pages} {133} (\bibinfo {year} {1995})}\BibitemShut
  {NoStop}%
\bibitem [{\citenamefont {Blais}\ \emph {et~al.}(2004)\citenamefont {Blais},
  \citenamefont {Huang}, \citenamefont {Wallraff}, \citenamefont {Girvin},\
  and\ \citenamefont {Schoelkopf}}]{blais2004cavity}%
  \BibitemOpen
  \bibfield  {author} {\bibinfo {author} {\bibfnamefont {A.}~\bibnamefont
  {Blais}}, \bibinfo {author} {\bibfnamefont {R.-S.}\ \bibnamefont {Huang}},
  \bibinfo {author} {\bibfnamefont {A.}~\bibnamefont {Wallraff}}, \bibinfo
  {author} {\bibfnamefont {S.~M.}\ \bibnamefont {Girvin}},\ and\ \bibinfo
  {author} {\bibfnamefont {R.~J.}\ \bibnamefont {Schoelkopf}},\ }\bibfield
  {title} {\bibinfo {title} {Cavity quantum electrodynamics for superconducting
  electrical circuits: An architecture for quantum computation},\ }\href@noop
  {} {\bibfield  {journal} {\bibinfo  {journal} {Physical Review A—Atomic,
  Molecular, and Optical Physics}\ }\textbf {\bibinfo {volume} {69}},\ \bibinfo
  {pages} {062320} (\bibinfo {year} {2004})}\BibitemShut {NoStop}%
\bibitem [{\citenamefont {Bourassa}\ \emph {et~al.}(2012)\citenamefont
  {Bourassa}, \citenamefont {Beaudoin}, \citenamefont {Gambetta},\ and\
  \citenamefont {Blais}}]{bourassa2012josephson}%
  \BibitemOpen
  \bibfield  {author} {\bibinfo {author} {\bibfnamefont {J.}~\bibnamefont
  {Bourassa}}, \bibinfo {author} {\bibfnamefont {F.}~\bibnamefont {Beaudoin}},
  \bibinfo {author} {\bibfnamefont {J.~M.}\ \bibnamefont {Gambetta}},\ and\
  \bibinfo {author} {\bibfnamefont {A.}~\bibnamefont {Blais}},\ }\bibfield
  {title} {\bibinfo {title} {Josephson-junction-embedded transmission-line
  resonators: From kerr medium to in-line transmon},\ }\href@noop {} {\bibfield
   {journal} {\bibinfo  {journal} {Physical Review A—Atomic, Molecular, and
  Optical Physics}\ }\textbf {\bibinfo {volume} {86}},\ \bibinfo {pages}
  {013814} (\bibinfo {year} {2012})}\BibitemShut {NoStop}%
\bibitem [{\citenamefont {Leib}\ \emph {et~al.}(2012)\citenamefont {Leib},
  \citenamefont {Deppe}, \citenamefont {Marx}, \citenamefont {Gross},\ and\
  \citenamefont {Hartmann}}]{leib2012networks}%
  \BibitemOpen
  \bibfield  {author} {\bibinfo {author} {\bibfnamefont {M.}~\bibnamefont
  {Leib}}, \bibinfo {author} {\bibfnamefont {F.}~\bibnamefont {Deppe}},
  \bibinfo {author} {\bibfnamefont {A.}~\bibnamefont {Marx}}, \bibinfo {author}
  {\bibfnamefont {R.}~\bibnamefont {Gross}},\ and\ \bibinfo {author}
  {\bibfnamefont {M.~J.}\ \bibnamefont {Hartmann}},\ }\bibfield  {title}
  {\bibinfo {title} {Networks of nonlinear superconducting transmission line
  resonators},\ }\href@noop {} {\bibfield  {journal} {\bibinfo  {journal} {New
  Journal of Physics}\ }\textbf {\bibinfo {volume} {14}},\ \bibinfo {pages}
  {075024} (\bibinfo {year} {2012})}\BibitemShut {NoStop}%
\bibitem [{\citenamefont {Mortensen}\ \emph {et~al.}(2016)\citenamefont
  {Mortensen}, \citenamefont {M{\o}lmer},\ and\ \citenamefont
  {Andersen}}]{mortensen2016normal}%
  \BibitemOpen
  \bibfield  {author} {\bibinfo {author} {\bibfnamefont {H.~L.}\ \bibnamefont
  {Mortensen}}, \bibinfo {author} {\bibfnamefont {K.}~\bibnamefont
  {M{\o}lmer}},\ and\ \bibinfo {author} {\bibfnamefont {C.~K.}\ \bibnamefont
  {Andersen}},\ }\bibfield  {title} {\bibinfo {title} {Normal modes of a
  superconducting transmission-line resonator with embedded lumped element
  circuit components},\ }\href@noop {} {\bibfield  {journal} {\bibinfo
  {journal} {Physical Review A}\ }\textbf {\bibinfo {volume} {94}},\ \bibinfo
  {pages} {053817} (\bibinfo {year} {2016})}\BibitemShut {NoStop}%
\bibitem [{\citenamefont {Parra-Rodriguez}\ \emph {et~al.}(2018)\citenamefont
  {Parra-Rodriguez}, \citenamefont {Rico}, \citenamefont {Solano},\ and\
  \citenamefont {Egusquiza}}]{parra2018quantum}%
  \BibitemOpen
  \bibfield  {author} {\bibinfo {author} {\bibfnamefont {A.}~\bibnamefont
  {Parra-Rodriguez}}, \bibinfo {author} {\bibfnamefont {E.}~\bibnamefont
  {Rico}}, \bibinfo {author} {\bibfnamefont {E.}~\bibnamefont {Solano}},\ and\
  \bibinfo {author} {\bibfnamefont {I.}~\bibnamefont {Egusquiza}},\ }\bibfield
  {title} {\bibinfo {title} {Quantum networks in divergence-free circuit qed},\
  }\href@noop {} {\bibfield  {journal} {\bibinfo  {journal} {Quantum Science
  and Technology}\ }\textbf {\bibinfo {volume} {3}},\ \bibinfo {pages} {024012}
  (\bibinfo {year} {2018})}\BibitemShut {NoStop}%
\bibitem [{\citenamefont {Minev}\ \emph
  {et~al.}(2021{\natexlab{a}})\citenamefont {Minev}, \citenamefont {McConkey},
  \citenamefont {Takita}, \citenamefont {Corcoles},\ and\ \citenamefont
  {Gambetta}}]{minev2021circuit}%
  \BibitemOpen
  \bibfield  {author} {\bibinfo {author} {\bibfnamefont {Z.~K.}\ \bibnamefont
  {Minev}}, \bibinfo {author} {\bibfnamefont {T.~G.}\ \bibnamefont {McConkey}},
  \bibinfo {author} {\bibfnamefont {M.}~\bibnamefont {Takita}}, \bibinfo
  {author} {\bibfnamefont {A.~D.}\ \bibnamefont {Corcoles}},\ and\ \bibinfo
  {author} {\bibfnamefont {J.~M.}\ \bibnamefont {Gambetta}},\ }\bibfield
  {title} {\bibinfo {title} {Circuit quantum electrodynamics (cqed) with
  modular quasi-lumped models},\ }\href@noop {} {\bibfield  {journal} {\bibinfo
   {journal} {arXiv preprint arXiv:2103.10344}\ } (\bibinfo {year}
  {2021}{\natexlab{a}})}\BibitemShut {NoStop}%
\bibitem [{\citenamefont {Egusquiza}\ and\ \citenamefont
  {Parra-Rodriguez}(2022)}]{egusquiza2022algebraic}%
  \BibitemOpen
  \bibfield  {author} {\bibinfo {author} {\bibfnamefont {I.}~\bibnamefont
  {Egusquiza}}\ and\ \bibinfo {author} {\bibfnamefont {A.}~\bibnamefont
  {Parra-Rodriguez}},\ }\bibfield  {title} {\bibinfo {title} {Algebraic
  canonical quantization of lumped superconducting networks},\ }\href@noop {}
  {\bibfield  {journal} {\bibinfo  {journal} {Physical Review B}\ }\textbf
  {\bibinfo {volume} {106}},\ \bibinfo {pages} {024510} (\bibinfo {year}
  {2022})}\BibitemShut {NoStop}%
\bibitem [{\citenamefont {Nigg}\ \emph {et~al.}(2012)\citenamefont {Nigg},
  \citenamefont {Paik}, \citenamefont {Vlastakis}, \citenamefont {Kirchmair},
  \citenamefont {Shankar}, \citenamefont {Frunzio}, \citenamefont {Devoret},
  \citenamefont {Schoelkopf},\ and\ \citenamefont {Girvin}}]{Nigg_2012}%
  \BibitemOpen
  \bibfield  {author} {\bibinfo {author} {\bibfnamefont {S.~E.}\ \bibnamefont
  {Nigg}}, \bibinfo {author} {\bibfnamefont {H.}~\bibnamefont {Paik}}, \bibinfo
  {author} {\bibfnamefont {B.}~\bibnamefont {Vlastakis}}, \bibinfo {author}
  {\bibfnamefont {G.}~\bibnamefont {Kirchmair}}, \bibinfo {author}
  {\bibfnamefont {S.}~\bibnamefont {Shankar}}, \bibinfo {author} {\bibfnamefont
  {L.}~\bibnamefont {Frunzio}}, \bibinfo {author} {\bibfnamefont {M.~H.}\
  \bibnamefont {Devoret}}, \bibinfo {author} {\bibfnamefont {R.~J.}\
  \bibnamefont {Schoelkopf}},\ and\ \bibinfo {author} {\bibfnamefont {S.~M.}\
  \bibnamefont {Girvin}},\ }\bibfield  {title} {\bibinfo {title} {Black-box
  superconducting circuit quantization},\ }\bibfield  {journal} {\bibinfo
  {journal} {Physical Review Letters}\ }\textbf {\bibinfo {volume} {108}},\
  \href {https://doi.org/10.1103/physrevlett.108.240502}
  {10.1103/physrevlett.108.240502} (\bibinfo {year} {2012})\BibitemShut
  {NoStop}%
\bibitem [{\citenamefont {Solgun}\ \emph {et~al.}(2014)\citenamefont {Solgun},
  \citenamefont {Abraham},\ and\ \citenamefont
  {DiVincenzo}}]{solgun2014blackbox}%
  \BibitemOpen
  \bibfield  {author} {\bibinfo {author} {\bibfnamefont {F.}~\bibnamefont
  {Solgun}}, \bibinfo {author} {\bibfnamefont {D.~W.}\ \bibnamefont
  {Abraham}},\ and\ \bibinfo {author} {\bibfnamefont {D.~P.}\ \bibnamefont
  {DiVincenzo}},\ }\bibfield  {title} {\bibinfo {title} {Blackbox quantization
  of superconducting circuits using exact impedance synthesis},\ }\href@noop {}
  {\bibfield  {journal} {\bibinfo  {journal} {Physical Review B}\ }\textbf
  {\bibinfo {volume} {90}},\ \bibinfo {pages} {134504} (\bibinfo {year}
  {2014})}\BibitemShut {NoStop}%
\bibitem [{\citenamefont {Minev}\ \emph
  {et~al.}(2021{\natexlab{b}})\citenamefont {Minev}, \citenamefont {Leghtas},
  \citenamefont {Mundhada}, \citenamefont {Christakis}, \citenamefont {Pop},\
  and\ \citenamefont {Devoret}}]{minev2021energy}%
  \BibitemOpen
  \bibfield  {author} {\bibinfo {author} {\bibfnamefont {Z.~K.}\ \bibnamefont
  {Minev}}, \bibinfo {author} {\bibfnamefont {Z.}~\bibnamefont {Leghtas}},
  \bibinfo {author} {\bibfnamefont {S.~O.}\ \bibnamefont {Mundhada}}, \bibinfo
  {author} {\bibfnamefont {L.}~\bibnamefont {Christakis}}, \bibinfo {author}
  {\bibfnamefont {I.~M.}\ \bibnamefont {Pop}},\ and\ \bibinfo {author}
  {\bibfnamefont {M.~H.}\ \bibnamefont {Devoret}},\ }\bibfield  {title}
  {\bibinfo {title} {Energy-participation quantization of josephson circuits},\
  }\href@noop {} {\bibfield  {journal} {\bibinfo  {journal} {npj Quantum
  Information}\ }\textbf {\bibinfo {volume} {7}},\ \bibinfo {pages} {131}
  (\bibinfo {year} {2021}{\natexlab{b}})}\BibitemShut {NoStop}%
\bibitem [{\citenamefont {Koch}\ \emph {et~al.}(2007)\citenamefont {Koch},
  \citenamefont {Terri}, \citenamefont {Gambetta}, \citenamefont {Houck},
  \citenamefont {Schuster}, \citenamefont {Majer}, \citenamefont {Blais},
  \citenamefont {Devoret}, \citenamefont {Girvin},\ and\ \citenamefont
  {Schoelkopf}}]{koch2007charge}%
  \BibitemOpen
  \bibfield  {author} {\bibinfo {author} {\bibfnamefont {J.}~\bibnamefont
  {Koch}}, \bibinfo {author} {\bibfnamefont {M.~Y.}\ \bibnamefont {Terri}},
  \bibinfo {author} {\bibfnamefont {J.}~\bibnamefont {Gambetta}}, \bibinfo
  {author} {\bibfnamefont {A.~A.}\ \bibnamefont {Houck}}, \bibinfo {author}
  {\bibfnamefont {D.~I.}\ \bibnamefont {Schuster}}, \bibinfo {author}
  {\bibfnamefont {J.}~\bibnamefont {Majer}}, \bibinfo {author} {\bibfnamefont
  {A.}~\bibnamefont {Blais}}, \bibinfo {author} {\bibfnamefont {M.~H.}\
  \bibnamefont {Devoret}}, \bibinfo {author} {\bibfnamefont {S.~M.}\
  \bibnamefont {Girvin}},\ and\ \bibinfo {author} {\bibfnamefont {R.~J.}\
  \bibnamefont {Schoelkopf}},\ }\bibfield  {title} {\bibinfo {title}
  {Charge-insensitive qubit design derived from the cooper pair box},\
  }\href@noop {} {\bibfield  {journal} {\bibinfo  {journal} {Physical Review
  A}\ }\textbf {\bibinfo {volume} {76}},\ \bibinfo {pages} {042319} (\bibinfo
  {year} {2007})}\BibitemShut {NoStop}%
\bibitem [{\citenamefont {Manucharyan}\ \emph {et~al.}(2009)\citenamefont
  {Manucharyan}, \citenamefont {Koch}, \citenamefont {Glazman},\ and\
  \citenamefont {Devoret}}]{manucharyan2009fluxonium}%
  \BibitemOpen
  \bibfield  {author} {\bibinfo {author} {\bibfnamefont {V.~E.}\ \bibnamefont
  {Manucharyan}}, \bibinfo {author} {\bibfnamefont {J.}~\bibnamefont {Koch}},
  \bibinfo {author} {\bibfnamefont {L.~I.}\ \bibnamefont {Glazman}},\ and\
  \bibinfo {author} {\bibfnamefont {M.~H.}\ \bibnamefont {Devoret}},\
  }\bibfield  {title} {\bibinfo {title} {Fluxonium: Single cooper-pair circuit
  free of charge offsets},\ }\href@noop {} {\bibfield  {journal} {\bibinfo
  {journal} {Science}\ }\textbf {\bibinfo {volume} {326}},\ \bibinfo {pages}
  {113} (\bibinfo {year} {2009})}\BibitemShut {NoStop}%
\bibitem [{\citenamefont {Nguyen}\ \emph {et~al.}(2019)\citenamefont {Nguyen},
  \citenamefont {Lin}, \citenamefont {Somoroff}, \citenamefont {Mencia},
  \citenamefont {Grabon},\ and\ \citenamefont {Manucharyan}}]{nguyen2019high}%
  \BibitemOpen
  \bibfield  {author} {\bibinfo {author} {\bibfnamefont {L.~B.}\ \bibnamefont
  {Nguyen}}, \bibinfo {author} {\bibfnamefont {Y.-H.}\ \bibnamefont {Lin}},
  \bibinfo {author} {\bibfnamefont {A.}~\bibnamefont {Somoroff}}, \bibinfo
  {author} {\bibfnamefont {R.}~\bibnamefont {Mencia}}, \bibinfo {author}
  {\bibfnamefont {N.}~\bibnamefont {Grabon}},\ and\ \bibinfo {author}
  {\bibfnamefont {V.~E.}\ \bibnamefont {Manucharyan}},\ }\bibfield  {title}
  {\bibinfo {title} {High-coherence fluxonium qubit},\ }\href@noop {}
  {\bibfield  {journal} {\bibinfo  {journal} {Physical Review X}\ }\textbf
  {\bibinfo {volume} {9}},\ \bibinfo {pages} {041041} (\bibinfo {year}
  {2019})}\BibitemShut {NoStop}%
\bibitem [{\citenamefont {Ding}\ \emph {et~al.}(2023)\citenamefont {Ding},
  \citenamefont {Hays}, \citenamefont {Sung}, \citenamefont {Kannan},
  \citenamefont {An}, \citenamefont {Di~Paolo}, \citenamefont {Karamlou},
  \citenamefont {Hazard}, \citenamefont {Azar}, \citenamefont {Kim} \emph
  {et~al.}}]{ding2023high}%
  \BibitemOpen
  \bibfield  {author} {\bibinfo {author} {\bibfnamefont {L.}~\bibnamefont
  {Ding}}, \bibinfo {author} {\bibfnamefont {M.}~\bibnamefont {Hays}}, \bibinfo
  {author} {\bibfnamefont {Y.}~\bibnamefont {Sung}}, \bibinfo {author}
  {\bibfnamefont {B.}~\bibnamefont {Kannan}}, \bibinfo {author} {\bibfnamefont
  {J.}~\bibnamefont {An}}, \bibinfo {author} {\bibfnamefont {A.}~\bibnamefont
  {Di~Paolo}}, \bibinfo {author} {\bibfnamefont {A.~H.}\ \bibnamefont
  {Karamlou}}, \bibinfo {author} {\bibfnamefont {T.~M.}\ \bibnamefont
  {Hazard}}, \bibinfo {author} {\bibfnamefont {K.}~\bibnamefont {Azar}},
  \bibinfo {author} {\bibfnamefont {D.~K.}\ \bibnamefont {Kim}}, \emph
  {et~al.},\ }\bibfield  {title} {\bibinfo {title} {High-fidelity,
  frequency-flexible two-qubit fluxonium gates with a transmon coupler},\
  }\href@noop {} {\bibfield  {journal} {\bibinfo  {journal} {Physical Review
  X}\ }\textbf {\bibinfo {volume} {13}},\ \bibinfo {pages} {031035} (\bibinfo
  {year} {2023})}\BibitemShut {NoStop}%
\bibitem [{\citenamefont {Somoroff}\ \emph {et~al.}(2023)\citenamefont
  {Somoroff}, \citenamefont {Ficheux}, \citenamefont {Mencia}, \citenamefont
  {Xiong}, \citenamefont {Kuzmin},\ and\ \citenamefont
  {Manucharyan}}]{somoroff2023millisecond}%
  \BibitemOpen
  \bibfield  {author} {\bibinfo {author} {\bibfnamefont {A.}~\bibnamefont
  {Somoroff}}, \bibinfo {author} {\bibfnamefont {Q.}~\bibnamefont {Ficheux}},
  \bibinfo {author} {\bibfnamefont {R.~A.}\ \bibnamefont {Mencia}}, \bibinfo
  {author} {\bibfnamefont {H.}~\bibnamefont {Xiong}}, \bibinfo {author}
  {\bibfnamefont {R.}~\bibnamefont {Kuzmin}},\ and\ \bibinfo {author}
  {\bibfnamefont {V.~E.}\ \bibnamefont {Manucharyan}},\ }\bibfield  {title}
  {\bibinfo {title} {Millisecond coherence in a superconducting qubit},\
  }\href@noop {} {\bibfield  {journal} {\bibinfo  {journal} {Physical Review
  Letters}\ }\textbf {\bibinfo {volume} {130}},\ \bibinfo {pages} {267001}
  (\bibinfo {year} {2023})}\BibitemShut {NoStop}%
\bibitem [{\citenamefont {Zhang}\ \emph {et~al.}(2024)\citenamefont {Zhang},
  \citenamefont {Ding}, \citenamefont {Weiss}, \citenamefont {Huang},
  \citenamefont {Ma}, \citenamefont {Guinn}, \citenamefont {Sussman},
  \citenamefont {Chitta}, \citenamefont {Chen}, \citenamefont {Houck} \emph
  {et~al.}}]{zhang2024tunable}%
  \BibitemOpen
  \bibfield  {author} {\bibinfo {author} {\bibfnamefont {H.}~\bibnamefont
  {Zhang}}, \bibinfo {author} {\bibfnamefont {C.}~\bibnamefont {Ding}},
  \bibinfo {author} {\bibfnamefont {D.}~\bibnamefont {Weiss}}, \bibinfo
  {author} {\bibfnamefont {Z.}~\bibnamefont {Huang}}, \bibinfo {author}
  {\bibfnamefont {Y.}~\bibnamefont {Ma}}, \bibinfo {author} {\bibfnamefont
  {C.}~\bibnamefont {Guinn}}, \bibinfo {author} {\bibfnamefont
  {S.}~\bibnamefont {Sussman}}, \bibinfo {author} {\bibfnamefont {S.~P.}\
  \bibnamefont {Chitta}}, \bibinfo {author} {\bibfnamefont {D.}~\bibnamefont
  {Chen}}, \bibinfo {author} {\bibfnamefont {A.~A.}\ \bibnamefont {Houck}},
  \emph {et~al.},\ }\bibfield  {title} {\bibinfo {title} {Tunable inductive
  coupler for high-fidelity gates between fluxonium qubits},\ }\href@noop {}
  {\bibfield  {journal} {\bibinfo  {journal} {PRX Quantum}\ }\textbf {\bibinfo
  {volume} {5}},\ \bibinfo {pages} {020326} (\bibinfo {year}
  {2024})}\BibitemShut {NoStop}%
\bibitem [{\citenamefont {Wang}\ \emph {et~al.}(2024)\citenamefont {Wang},
  \citenamefont {Lu}, \citenamefont {Zhan}, \citenamefont {Ma}, \citenamefont
  {Wu}, \citenamefont {Sun}, \citenamefont {Deng}, \citenamefont {Bai},
  \citenamefont {Bao}, \citenamefont {Chang} \emph
  {et~al.}}]{wang2024achieving}%
  \BibitemOpen
  \bibfield  {author} {\bibinfo {author} {\bibfnamefont {F.}~\bibnamefont
  {Wang}}, \bibinfo {author} {\bibfnamefont {K.}~\bibnamefont {Lu}}, \bibinfo
  {author} {\bibfnamefont {H.}~\bibnamefont {Zhan}}, \bibinfo {author}
  {\bibfnamefont {L.}~\bibnamefont {Ma}}, \bibinfo {author} {\bibfnamefont
  {F.}~\bibnamefont {Wu}}, \bibinfo {author} {\bibfnamefont {H.}~\bibnamefont
  {Sun}}, \bibinfo {author} {\bibfnamefont {H.}~\bibnamefont {Deng}}, \bibinfo
  {author} {\bibfnamefont {Y.}~\bibnamefont {Bai}}, \bibinfo {author}
  {\bibfnamefont {F.}~\bibnamefont {Bao}}, \bibinfo {author} {\bibfnamefont
  {X.}~\bibnamefont {Chang}}, \emph {et~al.},\ }\bibfield  {title} {\bibinfo
  {title} {Achieving millisecond coherence fluxonium through overlap josephson
  junctions},\ }\href@noop {} {\bibfield  {journal} {\bibinfo  {journal} {arXiv
  preprint arXiv:2405.05481}\ } (\bibinfo {year} {2024})}\BibitemShut {NoStop}%
\bibitem [{\citenamefont {Lin}\ \emph {et~al.}(2018)\citenamefont {Lin},
  \citenamefont {Nguyen}, \citenamefont {Grabon}, \citenamefont {San~Miguel},
  \citenamefont {Pankratova},\ and\ \citenamefont
  {Manucharyan}}]{lin2018demonstration}%
  \BibitemOpen
  \bibfield  {author} {\bibinfo {author} {\bibfnamefont {Y.-H.}\ \bibnamefont
  {Lin}}, \bibinfo {author} {\bibfnamefont {L.~B.}\ \bibnamefont {Nguyen}},
  \bibinfo {author} {\bibfnamefont {N.}~\bibnamefont {Grabon}}, \bibinfo
  {author} {\bibfnamefont {J.}~\bibnamefont {San~Miguel}}, \bibinfo {author}
  {\bibfnamefont {N.}~\bibnamefont {Pankratova}},\ and\ \bibinfo {author}
  {\bibfnamefont {V.~E.}\ \bibnamefont {Manucharyan}},\ }\bibfield  {title}
  {\bibinfo {title} {Demonstration of protection of a superconducting qubit
  from energy decay},\ }\href@noop {} {\bibfield  {journal} {\bibinfo
  {journal} {Physical review letters}\ }\textbf {\bibinfo {volume} {120}},\
  \bibinfo {pages} {150503} (\bibinfo {year} {2018})}\BibitemShut {NoStop}%
\bibitem [{\citenamefont {Earnest}\ \emph {et~al.}(2018)\citenamefont
  {Earnest}, \citenamefont {Chakram}, \citenamefont {Lu}, \citenamefont
  {Irons}, \citenamefont {Naik}, \citenamefont {Leung}, \citenamefont {Ocola},
  \citenamefont {Czaplewski}, \citenamefont {Baker}, \citenamefont {Lawrence}
  \emph {et~al.}}]{earnest2018realization}%
  \BibitemOpen
  \bibfield  {author} {\bibinfo {author} {\bibfnamefont {N.}~\bibnamefont
  {Earnest}}, \bibinfo {author} {\bibfnamefont {S.}~\bibnamefont {Chakram}},
  \bibinfo {author} {\bibfnamefont {Y.}~\bibnamefont {Lu}}, \bibinfo {author}
  {\bibfnamefont {N.}~\bibnamefont {Irons}}, \bibinfo {author} {\bibfnamefont
  {R.~K.}\ \bibnamefont {Naik}}, \bibinfo {author} {\bibfnamefont
  {N.}~\bibnamefont {Leung}}, \bibinfo {author} {\bibfnamefont
  {L.}~\bibnamefont {Ocola}}, \bibinfo {author} {\bibfnamefont {D.~A.}\
  \bibnamefont {Czaplewski}}, \bibinfo {author} {\bibfnamefont
  {B.}~\bibnamefont {Baker}}, \bibinfo {author} {\bibfnamefont
  {J.}~\bibnamefont {Lawrence}}, \emph {et~al.},\ }\bibfield  {title} {\bibinfo
  {title} {Realization of a $\lambda$ system with metastable states of a
  capacitively shunted fluxonium},\ }\href@noop {} {\bibfield  {journal}
  {\bibinfo  {journal} {Physical review letters}\ }\textbf {\bibinfo {volume}
  {120}},\ \bibinfo {pages} {150504} (\bibinfo {year} {2018})}\BibitemShut
  {NoStop}%
\bibitem [{\citenamefont {Josephson}(1962)}]{josephson1962possible}%
  \BibitemOpen
  \bibfield  {author} {\bibinfo {author} {\bibfnamefont {B.~D.}\ \bibnamefont
  {Josephson}},\ }\bibfield  {title} {\bibinfo {title} {Possible new effects in
  superconductive tunnelling},\ }\href@noop {} {\bibfield  {journal} {\bibinfo
  {journal} {Physics letters}\ }\textbf {\bibinfo {volume} {1}},\ \bibinfo
  {pages} {251} (\bibinfo {year} {1962})}\BibitemShut {NoStop}%
\bibitem [{\citenamefont {Josephson}(1965)}]{josephson1965phys}%
  \BibitemOpen
  \bibfield  {author} {\bibinfo {author} {\bibfnamefont {B.}~\bibnamefont
  {Josephson}},\ }\bibfield  {title} {\bibinfo {title} {Phys. letters 1, 251
  (1962)},\ }\href@noop {} {\bibfield  {journal} {\bibinfo  {journal} {Advan.
  Phys}\ }\textbf {\bibinfo {volume} {14}},\ \bibinfo {pages} {419} (\bibinfo
  {year} {1965})}\BibitemShut {NoStop}%
\bibitem [{\citenamefont {Masluk}\ \emph {et~al.}(2012)\citenamefont {Masluk},
  \citenamefont {Pop}, \citenamefont {Kamal}, \citenamefont {Minev},\ and\
  \citenamefont {Devoret}}]{masluk2012microwave}%
  \BibitemOpen
  \bibfield  {author} {\bibinfo {author} {\bibfnamefont {N.~A.}\ \bibnamefont
  {Masluk}}, \bibinfo {author} {\bibfnamefont {I.~M.}\ \bibnamefont {Pop}},
  \bibinfo {author} {\bibfnamefont {A.}~\bibnamefont {Kamal}}, \bibinfo
  {author} {\bibfnamefont {Z.~K.}\ \bibnamefont {Minev}},\ and\ \bibinfo
  {author} {\bibfnamefont {M.~H.}\ \bibnamefont {Devoret}},\ }\bibfield
  {title} {\bibinfo {title} {Microwave characterization of josephson junction
  arrays: Implementing<? format?> a low loss superinductance},\ }\href@noop {}
  {\bibfield  {journal} {\bibinfo  {journal} {Physical review letters}\
  }\textbf {\bibinfo {volume} {109}},\ \bibinfo {pages} {137002} (\bibinfo
  {year} {2012})}\BibitemShut {NoStop}%
\bibitem [{\citenamefont {Puertas~Mart{\'\i}nez}\ \emph
  {et~al.}(2019)\citenamefont {Puertas~Mart{\'\i}nez}, \citenamefont
  {L{\'e}ger}, \citenamefont {Gheeraert}, \citenamefont {Dassonneville},
  \citenamefont {Planat}, \citenamefont {Foroughi}, \citenamefont {Krupko},
  \citenamefont {Buisson}, \citenamefont {Naud}, \citenamefont {Hasch-Guichard}
  \emph {et~al.}}]{puertas2019tunable}%
  \BibitemOpen
  \bibfield  {author} {\bibinfo {author} {\bibfnamefont {J.}~\bibnamefont
  {Puertas~Mart{\'\i}nez}}, \bibinfo {author} {\bibfnamefont {S.}~\bibnamefont
  {L{\'e}ger}}, \bibinfo {author} {\bibfnamefont {N.}~\bibnamefont
  {Gheeraert}}, \bibinfo {author} {\bibfnamefont {R.}~\bibnamefont
  {Dassonneville}}, \bibinfo {author} {\bibfnamefont {L.}~\bibnamefont
  {Planat}}, \bibinfo {author} {\bibfnamefont {F.}~\bibnamefont {Foroughi}},
  \bibinfo {author} {\bibfnamefont {Y.}~\bibnamefont {Krupko}}, \bibinfo
  {author} {\bibfnamefont {O.}~\bibnamefont {Buisson}}, \bibinfo {author}
  {\bibfnamefont {C.}~\bibnamefont {Naud}}, \bibinfo {author} {\bibfnamefont
  {W.}~\bibnamefont {Hasch-Guichard}}, \emph {et~al.},\ }\bibfield  {title}
  {\bibinfo {title} {A tunable josephson platform to explore many-body quantum
  optics in circuit-qed},\ }\href@noop {} {\bibfield  {journal} {\bibinfo
  {journal} {npj Quantum Information}\ }\textbf {\bibinfo {volume} {5}},\
  \bibinfo {pages} {19} (\bibinfo {year} {2019})}\BibitemShut {NoStop}%
\bibitem [{\citenamefont {Hazard}\ \emph {et~al.}(2019)\citenamefont {Hazard},
  \citenamefont {Gyenis}, \citenamefont {Di~Paolo}, \citenamefont {Asfaw},
  \citenamefont {Lyon}, \citenamefont {Blais},\ and\ \citenamefont
  {Houck}}]{hazard2019nanowire}%
  \BibitemOpen
  \bibfield  {author} {\bibinfo {author} {\bibfnamefont {T.}~\bibnamefont
  {Hazard}}, \bibinfo {author} {\bibfnamefont {A.}~\bibnamefont {Gyenis}},
  \bibinfo {author} {\bibfnamefont {A.}~\bibnamefont {Di~Paolo}}, \bibinfo
  {author} {\bibfnamefont {A.}~\bibnamefont {Asfaw}}, \bibinfo {author}
  {\bibfnamefont {S.}~\bibnamefont {Lyon}}, \bibinfo {author} {\bibfnamefont
  {A.}~\bibnamefont {Blais}},\ and\ \bibinfo {author} {\bibfnamefont
  {A.}~\bibnamefont {Houck}},\ }\bibfield  {title} {\bibinfo {title} {Nanowire
  superinductance fluxonium qubit},\ }\href@noop {} {\bibfield  {journal}
  {\bibinfo  {journal} {Physical review letters}\ }\textbf {\bibinfo {volume}
  {122}},\ \bibinfo {pages} {010504} (\bibinfo {year} {2019})}\BibitemShut
  {NoStop}%
\bibitem [{\citenamefont {Niepce}\ \emph {et~al.}(2019)\citenamefont {Niepce},
  \citenamefont {Burnett},\ and\ \citenamefont {Bylander}}]{niepce2019high}%
  \BibitemOpen
  \bibfield  {author} {\bibinfo {author} {\bibfnamefont {D.}~\bibnamefont
  {Niepce}}, \bibinfo {author} {\bibfnamefont {J.}~\bibnamefont {Burnett}},\
  and\ \bibinfo {author} {\bibfnamefont {J.}~\bibnamefont {Bylander}},\
  }\bibfield  {title} {\bibinfo {title} {High kinetic inductance nb n nanowire
  superinductors},\ }\href@noop {} {\bibfield  {journal} {\bibinfo  {journal}
  {Physical Review Applied}\ }\textbf {\bibinfo {volume} {11}},\ \bibinfo
  {pages} {044014} (\bibinfo {year} {2019})}\BibitemShut {NoStop}%
\bibitem [{\citenamefont {Gr{\"u}nhaupt}\ \emph {et~al.}(2019)\citenamefont
  {Gr{\"u}nhaupt}, \citenamefont {Spiecker}, \citenamefont {Gusenkova},
  \citenamefont {Maleeva}, \citenamefont {Skacel}, \citenamefont {Takmakov},
  \citenamefont {Valenti}, \citenamefont {Winkel}, \citenamefont {Rotzinger},
  \citenamefont {Wernsdorfer} \emph {et~al.}}]{grunhaupt2019granular}%
  \BibitemOpen
  \bibfield  {author} {\bibinfo {author} {\bibfnamefont {L.}~\bibnamefont
  {Gr{\"u}nhaupt}}, \bibinfo {author} {\bibfnamefont {M.}~\bibnamefont
  {Spiecker}}, \bibinfo {author} {\bibfnamefont {D.}~\bibnamefont {Gusenkova}},
  \bibinfo {author} {\bibfnamefont {N.}~\bibnamefont {Maleeva}}, \bibinfo
  {author} {\bibfnamefont {S.~T.}\ \bibnamefont {Skacel}}, \bibinfo {author}
  {\bibfnamefont {I.}~\bibnamefont {Takmakov}}, \bibinfo {author}
  {\bibfnamefont {F.}~\bibnamefont {Valenti}}, \bibinfo {author} {\bibfnamefont
  {P.}~\bibnamefont {Winkel}}, \bibinfo {author} {\bibfnamefont
  {H.}~\bibnamefont {Rotzinger}}, \bibinfo {author} {\bibfnamefont
  {W.}~\bibnamefont {Wernsdorfer}}, \emph {et~al.},\ }\bibfield  {title}
  {\bibinfo {title} {Granular aluminium as a superconducting material for
  high-impedance quantum circuits},\ }\href@noop {} {\bibfield  {journal}
  {\bibinfo  {journal} {Nature materials}\ }\textbf {\bibinfo {volume} {18}},\
  \bibinfo {pages} {816} (\bibinfo {year} {2019})}\BibitemShut {NoStop}%
\bibitem [{\citenamefont {Rieger}\ \emph {et~al.}(2023)\citenamefont {Rieger},
  \citenamefont {G{\"u}nzler}, \citenamefont {Spiecker}, \citenamefont
  {Paluch}, \citenamefont {Winkel}, \citenamefont {Hahn}, \citenamefont
  {Hohmann}, \citenamefont {Bacher}, \citenamefont {Wernsdorfer},\ and\
  \citenamefont {Pop}}]{rieger2023granular}%
  \BibitemOpen
  \bibfield  {author} {\bibinfo {author} {\bibfnamefont {D.}~\bibnamefont
  {Rieger}}, \bibinfo {author} {\bibfnamefont {S.}~\bibnamefont {G{\"u}nzler}},
  \bibinfo {author} {\bibfnamefont {M.}~\bibnamefont {Spiecker}}, \bibinfo
  {author} {\bibfnamefont {P.}~\bibnamefont {Paluch}}, \bibinfo {author}
  {\bibfnamefont {P.}~\bibnamefont {Winkel}}, \bibinfo {author} {\bibfnamefont
  {L.}~\bibnamefont {Hahn}}, \bibinfo {author} {\bibfnamefont {J.}~\bibnamefont
  {Hohmann}}, \bibinfo {author} {\bibfnamefont {A.}~\bibnamefont {Bacher}},
  \bibinfo {author} {\bibfnamefont {W.}~\bibnamefont {Wernsdorfer}},\ and\
  \bibinfo {author} {\bibfnamefont {I.}~\bibnamefont {Pop}},\ }\bibfield
  {title} {\bibinfo {title} {Granular aluminium nanojunction fluxonium qubit},\
  }\href@noop {} {\bibfield  {journal} {\bibinfo  {journal} {Nature Materials}\
  }\textbf {\bibinfo {volume} {22}},\ \bibinfo {pages} {194} (\bibinfo {year}
  {2023})}\BibitemShut {NoStop}%
\bibitem [{\citenamefont {Stefanski}\ and\ \citenamefont
  {Andersen}(2024)}]{stefanski2024flux}%
  \BibitemOpen
  \bibfield  {author} {\bibinfo {author} {\bibfnamefont {T.~V.}\ \bibnamefont
  {Stefanski}}\ and\ \bibinfo {author} {\bibfnamefont {C.~K.}\ \bibnamefont
  {Andersen}},\ }\bibfield  {title} {\bibinfo {title} {Flux-pulse-assisted
  readout of a fluxonium qubit},\ }\href@noop {} {\bibfield  {journal}
  {\bibinfo  {journal} {Physical Review Applied}\ }\textbf {\bibinfo {volume}
  {22}},\ \bibinfo {pages} {014079} (\bibinfo {year} {2024})}\BibitemShut
  {NoStop}%
\bibitem [{\citenamefont {Lin}\ \emph {et~al.}(2024)\citenamefont {Lin},
  \citenamefont {Cho}, \citenamefont {Chen}, \citenamefont {Vavilov},
  \citenamefont {Wang},\ and\ \citenamefont {Manucharyan}}]{lin202424}%
  \BibitemOpen
  \bibfield  {author} {\bibinfo {author} {\bibfnamefont {W.-J.}\ \bibnamefont
  {Lin}}, \bibinfo {author} {\bibfnamefont {H.}~\bibnamefont {Cho}}, \bibinfo
  {author} {\bibfnamefont {Y.}~\bibnamefont {Chen}}, \bibinfo {author}
  {\bibfnamefont {M.~G.}\ \bibnamefont {Vavilov}}, \bibinfo {author}
  {\bibfnamefont {C.}~\bibnamefont {Wang}},\ and\ \bibinfo {author}
  {\bibfnamefont {V.~E.}\ \bibnamefont {Manucharyan}},\ }\bibfield  {title}
  {\bibinfo {title} {24 days-stable cnot-gate on fluxonium qubits with over
  99.9\% fidelity},\ }\href@noop {} {\bibfield  {journal} {\bibinfo  {journal}
  {arXiv preprint arXiv:2407.15783}\ } (\bibinfo {year} {2024})}\BibitemShut
  {NoStop}%
\bibitem [{\citenamefont {Zhu}\ \emph {et~al.}(2013)\citenamefont {Zhu},
  \citenamefont {Ferguson}, \citenamefont {Manucharyan},\ and\ \citenamefont
  {Koch}}]{zhu2013circuit}%
  \BibitemOpen
  \bibfield  {author} {\bibinfo {author} {\bibfnamefont {G.}~\bibnamefont
  {Zhu}}, \bibinfo {author} {\bibfnamefont {D.~G.}\ \bibnamefont {Ferguson}},
  \bibinfo {author} {\bibfnamefont {V.~E.}\ \bibnamefont {Manucharyan}},\ and\
  \bibinfo {author} {\bibfnamefont {J.}~\bibnamefont {Koch}},\ }\bibfield
  {title} {\bibinfo {title} {Circuit qed with fluxonium qubits: Theory of the
  dispersive regime},\ }\href@noop {} {\bibfield  {journal} {\bibinfo
  {journal} {Physical Review B—Condensed Matter and Materials Physics}\
  }\textbf {\bibinfo {volume} {87}},\ \bibinfo {pages} {024510} (\bibinfo
  {year} {2013})}\BibitemShut {NoStop}%
\bibitem [{\citenamefont {Chiaro}\ \emph {et~al.}(2016)\citenamefont {Chiaro},
  \citenamefont {Megrant}, \citenamefont {Dunsworth}, \citenamefont {Chen},
  \citenamefont {Barends}, \citenamefont {Campbell}, \citenamefont {Chen},
  \citenamefont {Fowler}, \citenamefont {Hoi}, \citenamefont {Jeffrey} \emph
  {et~al.}}]{chiaro2016dielectric}%
  \BibitemOpen
  \bibfield  {author} {\bibinfo {author} {\bibfnamefont {B.}~\bibnamefont
  {Chiaro}}, \bibinfo {author} {\bibfnamefont {A.}~\bibnamefont {Megrant}},
  \bibinfo {author} {\bibfnamefont {A.}~\bibnamefont {Dunsworth}}, \bibinfo
  {author} {\bibfnamefont {Z.}~\bibnamefont {Chen}}, \bibinfo {author}
  {\bibfnamefont {R.}~\bibnamefont {Barends}}, \bibinfo {author} {\bibfnamefont
  {B.}~\bibnamefont {Campbell}}, \bibinfo {author} {\bibfnamefont
  {Y.}~\bibnamefont {Chen}}, \bibinfo {author} {\bibfnamefont {A.}~\bibnamefont
  {Fowler}}, \bibinfo {author} {\bibfnamefont {I.}~\bibnamefont {Hoi}},
  \bibinfo {author} {\bibfnamefont {E.}~\bibnamefont {Jeffrey}}, \emph
  {et~al.},\ }\bibfield  {title} {\bibinfo {title} {Dielectric surface loss in
  superconducting resonators with flux-trapping holes},\ }\href@noop {}
  {\bibfield  {journal} {\bibinfo  {journal} {Superconductor Science and
  Technology}\ }\textbf {\bibinfo {volume} {29}},\ \bibinfo {pages} {104006}
  (\bibinfo {year} {2016})}\BibitemShut {NoStop}%
\bibitem [{\citenamefont {Potts}\ \emph {et~al.}(2001)\citenamefont {Potts},
  \citenamefont {Routley}, \citenamefont {Parker}, \citenamefont {Baumberg},\
  and\ \citenamefont {De~Groot}}]{potts2001novel}%
  \BibitemOpen
  \bibfield  {author} {\bibinfo {author} {\bibfnamefont {A.}~\bibnamefont
  {Potts}}, \bibinfo {author} {\bibfnamefont {P.}~\bibnamefont {Routley}},
  \bibinfo {author} {\bibfnamefont {G.~J.}\ \bibnamefont {Parker}}, \bibinfo
  {author} {\bibfnamefont {J.}~\bibnamefont {Baumberg}},\ and\ \bibinfo
  {author} {\bibfnamefont {P.}~\bibnamefont {De~Groot}},\ }\bibfield  {title}
  {\bibinfo {title} {Novel fabrication methods for submicrometer josephson
  junction qubits},\ }\href@noop {} {\bibfield  {journal} {\bibinfo  {journal}
  {Journal of Materials Science: Materials in Electronics}\ }\textbf {\bibinfo
  {volume} {12}},\ \bibinfo {pages} {289} (\bibinfo {year} {2001})}\BibitemShut
  {NoStop}%
\bibitem [{\citenamefont {Dunsworth}\ \emph {et~al.}(2017)\citenamefont
  {Dunsworth}, \citenamefont {Megrant}, \citenamefont {Quintana}, \citenamefont
  {Chen}, \citenamefont {Barends}, \citenamefont {Burkett}, \citenamefont
  {Foxen}, \citenamefont {Chen}, \citenamefont {Chiaro}, \citenamefont {Fowler}
  \emph {et~al.}}]{dunsworth2017characterization}%
  \BibitemOpen
  \bibfield  {author} {\bibinfo {author} {\bibfnamefont {A.}~\bibnamefont
  {Dunsworth}}, \bibinfo {author} {\bibfnamefont {A.}~\bibnamefont {Megrant}},
  \bibinfo {author} {\bibfnamefont {C.}~\bibnamefont {Quintana}}, \bibinfo
  {author} {\bibfnamefont {Z.}~\bibnamefont {Chen}}, \bibinfo {author}
  {\bibfnamefont {R.}~\bibnamefont {Barends}}, \bibinfo {author} {\bibfnamefont
  {B.}~\bibnamefont {Burkett}}, \bibinfo {author} {\bibfnamefont
  {B.}~\bibnamefont {Foxen}}, \bibinfo {author} {\bibfnamefont
  {Y.}~\bibnamefont {Chen}}, \bibinfo {author} {\bibfnamefont {B.}~\bibnamefont
  {Chiaro}}, \bibinfo {author} {\bibfnamefont {A.}~\bibnamefont {Fowler}},
  \emph {et~al.},\ }\bibfield  {title} {\bibinfo {title} {Characterization and
  reduction of capacitive loss induced by sub-micron josephson junction
  fabrication in superconducting qubits},\ }\href@noop {} {\bibfield  {journal}
  {\bibinfo  {journal} {Applied Physics Letters}\ }\textbf {\bibinfo {volume}
  {111}} (\bibinfo {year} {2017})}\BibitemShut {NoStop}%
\bibitem [{\citenamefont {Minev}\ \emph
  {et~al.}(2021{\natexlab{c}})\citenamefont {Minev}, \citenamefont {McConkey},
  \citenamefont {Drysdale}, \citenamefont {Shah}, \citenamefont {Wang},
  \citenamefont {Facchini}, \citenamefont {Harper}, \citenamefont {Blair},
  \citenamefont {Zhang}, \citenamefont {Lanzillo}, \citenamefont {Mukesh},
  \citenamefont {Shanks}, \citenamefont {Warren},\ and\ \citenamefont
  {Gambetta}}]{Qiskit_Metal}%
  \BibitemOpen
  \bibfield  {author} {\bibinfo {author} {\bibfnamefont {Z.~K.}\ \bibnamefont
  {Minev}}, \bibinfo {author} {\bibfnamefont {T.~G.}\ \bibnamefont {McConkey}},
  \bibinfo {author} {\bibfnamefont {J.}~\bibnamefont {Drysdale}}, \bibinfo
  {author} {\bibfnamefont {P.}~\bibnamefont {Shah}}, \bibinfo {author}
  {\bibfnamefont {D.}~\bibnamefont {Wang}}, \bibinfo {author} {\bibfnamefont
  {M.}~\bibnamefont {Facchini}}, \bibinfo {author} {\bibfnamefont
  {G.}~\bibnamefont {Harper}}, \bibinfo {author} {\bibfnamefont
  {J.}~\bibnamefont {Blair}}, \bibinfo {author} {\bibfnamefont
  {H.}~\bibnamefont {Zhang}}, \bibinfo {author} {\bibfnamefont
  {N.}~\bibnamefont {Lanzillo}}, \bibinfo {author} {\bibfnamefont
  {S.}~\bibnamefont {Mukesh}}, \bibinfo {author} {\bibfnamefont
  {W.}~\bibnamefont {Shanks}}, \bibinfo {author} {\bibfnamefont
  {C.}~\bibnamefont {Warren}},\ and\ \bibinfo {author} {\bibfnamefont {J.~M.}\
  \bibnamefont {Gambetta}},\ }\href {https://doi.org/10.5281/zenodo.4618153}
  {\bibinfo {title} {{Qiskit Metal: An Open-Source Framework for Quantum Device
  Design {\&} Analysis}}} (\bibinfo {year} {2021}{\natexlab{c}})\BibitemShut
  {NoStop}%
\bibitem [{\citenamefont {Deppe}\ \emph {et~al.}(2004)\citenamefont {Deppe},
  \citenamefont {Saito}, \citenamefont {Tanaka},\ and\ \citenamefont
  {Takayanagi}}]{deppe2004determination}%
  \BibitemOpen
  \bibfield  {author} {\bibinfo {author} {\bibfnamefont {F.}~\bibnamefont
  {Deppe}}, \bibinfo {author} {\bibfnamefont {S.}~\bibnamefont {Saito}},
  \bibinfo {author} {\bibfnamefont {H.}~\bibnamefont {Tanaka}},\ and\ \bibinfo
  {author} {\bibfnamefont {H.}~\bibnamefont {Takayanagi}},\ }\bibfield  {title}
  {\bibinfo {title} {Determination of the capacitance of nm scale josephson
  junctions},\ }\href@noop {} {\bibfield  {journal} {\bibinfo  {journal}
  {Journal of applied physics}\ }\textbf {\bibinfo {volume} {95}},\ \bibinfo
  {pages} {2607} (\bibinfo {year} {2004})}\BibitemShut {NoStop}%
\bibitem [{\citenamefont {Kroll}\ \emph {et~al.}(2019)\citenamefont {Kroll},
  \citenamefont {Borsoi}, \citenamefont {Van Der~Enden}, \citenamefont
  {Uilhoorn}, \citenamefont {De~Jong}, \citenamefont {Quintero-P{\'e}rez},
  \citenamefont {Van~Woerkom}, \citenamefont {Bruno}, \citenamefont {Plissard},
  \citenamefont {Car} \emph {et~al.}}]{kroll2019magnetic}%
  \BibitemOpen
  \bibfield  {author} {\bibinfo {author} {\bibfnamefont {J.~G.}\ \bibnamefont
  {Kroll}}, \bibinfo {author} {\bibfnamefont {F.}~\bibnamefont {Borsoi}},
  \bibinfo {author} {\bibfnamefont {K.}~\bibnamefont {Van Der~Enden}}, \bibinfo
  {author} {\bibfnamefont {W.}~\bibnamefont {Uilhoorn}}, \bibinfo {author}
  {\bibfnamefont {D.}~\bibnamefont {De~Jong}}, \bibinfo {author} {\bibfnamefont
  {M.}~\bibnamefont {Quintero-P{\'e}rez}}, \bibinfo {author} {\bibfnamefont
  {D.}~\bibnamefont {Van~Woerkom}}, \bibinfo {author} {\bibfnamefont
  {A.}~\bibnamefont {Bruno}}, \bibinfo {author} {\bibfnamefont
  {S.}~\bibnamefont {Plissard}}, \bibinfo {author} {\bibfnamefont
  {D.}~\bibnamefont {Car}}, \emph {et~al.},\ }\bibfield  {title} {\bibinfo
  {title} {Magnetic-field-resilient superconducting coplanar-waveguide
  resonators for hybrid circuit quantum electrodynamics experiments},\
  }\href@noop {} {\bibfield  {journal} {\bibinfo  {journal} {Physical Review
  Applied}\ }\textbf {\bibinfo {volume} {11}},\ \bibinfo {pages} {064053}
  (\bibinfo {year} {2019})}\BibitemShut {NoStop}%
\bibitem [{\citenamefont {Moskalenko}\ \emph {et~al.}(2022)\citenamefont
  {Moskalenko}, \citenamefont {Simakov}, \citenamefont {Abramov}, \citenamefont
  {Grigorev}, \citenamefont {Moskalev}, \citenamefont {Pishchimova},
  \citenamefont {Smirnov}, \citenamefont {Zikiy}, \citenamefont {Rodionov},\
  and\ \citenamefont {Besedin}}]{moskalenko2022high}%
  \BibitemOpen
  \bibfield  {author} {\bibinfo {author} {\bibfnamefont {I.~N.}\ \bibnamefont
  {Moskalenko}}, \bibinfo {author} {\bibfnamefont {I.~A.}\ \bibnamefont
  {Simakov}}, \bibinfo {author} {\bibfnamefont {N.~N.}\ \bibnamefont
  {Abramov}}, \bibinfo {author} {\bibfnamefont {A.~A.}\ \bibnamefont
  {Grigorev}}, \bibinfo {author} {\bibfnamefont {D.~O.}\ \bibnamefont
  {Moskalev}}, \bibinfo {author} {\bibfnamefont {A.~A.}\ \bibnamefont
  {Pishchimova}}, \bibinfo {author} {\bibfnamefont {N.~S.}\ \bibnamefont
  {Smirnov}}, \bibinfo {author} {\bibfnamefont {E.~V.}\ \bibnamefont {Zikiy}},
  \bibinfo {author} {\bibfnamefont {I.~A.}\ \bibnamefont {Rodionov}},\ and\
  \bibinfo {author} {\bibfnamefont {I.~S.}\ \bibnamefont {Besedin}},\
  }\bibfield  {title} {\bibinfo {title} {High fidelity two-qubit gates on
  fluxoniums using a tunable coupler},\ }\href@noop {} {\bibfield  {journal}
  {\bibinfo  {journal} {npj Quantum Information}\ }\textbf {\bibinfo {volume}
  {8}},\ \bibinfo {pages} {130} (\bibinfo {year} {2022})}\BibitemShut {NoStop}%
\bibitem [{\citenamefont {Chitta}\ \emph {et~al.}(2022)\citenamefont {Chitta},
  \citenamefont {Zhao}, \citenamefont {Huang}, \citenamefont {Mondragon-Shem},\
  and\ \citenamefont {Koch}}]{chitta2022computer}%
  \BibitemOpen
  \bibfield  {author} {\bibinfo {author} {\bibfnamefont {S.~P.}\ \bibnamefont
  {Chitta}}, \bibinfo {author} {\bibfnamefont {T.}~\bibnamefont {Zhao}},
  \bibinfo {author} {\bibfnamefont {Z.}~\bibnamefont {Huang}}, \bibinfo
  {author} {\bibfnamefont {I.}~\bibnamefont {Mondragon-Shem}},\ and\ \bibinfo
  {author} {\bibfnamefont {J.}~\bibnamefont {Koch}},\ }\bibfield  {title}
  {\bibinfo {title} {Computer-aided quantization and numerical analysis of
  superconducting circuits},\ }\href@noop {} {\bibfield  {journal} {\bibinfo
  {journal} {New Journal of Physics}\ }\textbf {\bibinfo {volume} {24}},\
  \bibinfo {pages} {103020} (\bibinfo {year} {2022})}\BibitemShut {NoStop}%
\bibitem [{\citenamefont {Groszkowski}\ and\ \citenamefont
  {Koch}(2021)}]{groszkowski2021scqubits}%
  \BibitemOpen
  \bibfield  {author} {\bibinfo {author} {\bibfnamefont {P.}~\bibnamefont
  {Groszkowski}}\ and\ \bibinfo {author} {\bibfnamefont {J.}~\bibnamefont
  {Koch}},\ }\bibfield  {title} {\bibinfo {title} {Scqubits: a python package
  for superconducting qubits},\ }\href@noop {} {\bibfield  {journal} {\bibinfo
  {journal} {Quantum}\ }\textbf {\bibinfo {volume} {5}},\ \bibinfo {pages}
  {583} (\bibinfo {year} {2021})}\BibitemShut {NoStop}%
\bibitem [{EPR()}]{EPR_Repo}%
  \BibitemOpen
  \href {https://github.com/AndersenQubitLab/andersen-lab-pyEPR.git} {}\bibinfo
  {note} {The extended EPR Repository:
  \url{https://github.com/AndersenQubitLab/andersen-lab-pyEPR.git}.}\BibitemShut
  {Stop}%
\bibitem [{Qis()}]{Qiskit-Metal_Repo}%
  \BibitemOpen
  \href {https://github.com/AndersenQubitLab/andersen-lab-qiskit-metal}
  {}\bibinfo {note} {Qiskit-Metal Repository:
  \url{https://github.com/AndersenQubitLab/andersen-lab-qiskit-metal}.}\BibitemShut
  {Stop}%
\bibitem [{Mea()}]{Measurement_Data_Repo}%
  \BibitemOpen
  \href {https://doi.org/10.4121/833e20fb-9863-4d0d-828a-d2ec44108e12}
  {}\bibinfo {note} {Measurement Data Repository:
  \url{https://doi.org/10.4121/833e20fb-9863-4d0d-828a-d2ec44108e12}.}\BibitemShut
  {Stop}%
\bibitem [{\citenamefont {Bruno}\ \emph {et~al.}(2015)\citenamefont {Bruno},
  \citenamefont {De~Lange}, \citenamefont {Asaad}, \citenamefont {Van
  Der~Enden}, \citenamefont {Langford},\ and\ \citenamefont
  {DiCarlo}}]{bruno2015reducing}%
  \BibitemOpen
  \bibfield  {author} {\bibinfo {author} {\bibfnamefont {A.}~\bibnamefont
  {Bruno}}, \bibinfo {author} {\bibfnamefont {G.}~\bibnamefont {De~Lange}},
  \bibinfo {author} {\bibfnamefont {S.}~\bibnamefont {Asaad}}, \bibinfo
  {author} {\bibfnamefont {K.}~\bibnamefont {Van Der~Enden}}, \bibinfo {author}
  {\bibfnamefont {N.}~\bibnamefont {Langford}},\ and\ \bibinfo {author}
  {\bibfnamefont {L.}~\bibnamefont {DiCarlo}},\ }\bibfield  {title} {\bibinfo
  {title} {Reducing intrinsic loss in superconducting resonators by surface
  treatment and deep etching of silicon substrates},\ }\href@noop {} {\bibfield
   {journal} {\bibinfo  {journal} {Applied Physics Letters}\ }\textbf {\bibinfo
  {volume} {106}} (\bibinfo {year} {2015})}\BibitemShut {NoStop}%
\end{thebibliography}%

\end{document}